\documentclass[11pt]{article}
\usepackage{jheppubmod}
\usepackage{cite}
\pdfoutput=1
\usepackage{cancel}
\usepackage{xcolor}
\usepackage{caption}  
\usepackage{cite}
\usepackage{relsize}
\usepackage{physics}
\usepackage{psfrag}
\usepackage{cancel}
\usepackage{array}
\usepackage{amssymb}
\usepackage{amsmath}
\usepackage[compat=1.1.0]{tikz-feynman}
\usepackage{amsthm}
\usepackage{float}
\usepackage{tikz}
\usetikzlibrary{decorations.markings}
\usepackage{tikz,lipsum,lmodern}
\usepackage[most]{tcolorbox}
\usepackage{hyperref}
\usepackage{xcolor}

\hypersetup{
	colorlinks=true,
	urlcolor=midnightblue,
	citecolor=purplemountainmajesty,
	linkcolor=rosebonbon,
}
\usepackage{color,hyperref}
\definecolor{mediumorchid}{rgb}{0.73, 0.33, 0.83}
\definecolor{twilightlavender}{rgb}{0.54, 0.29, 0.42}
\definecolor{purplemountainmajesty}{rgb}{0.59, 0.47, 0.71}
\definecolor{brightmaroon}{rgb}{0.76, 0.13, 0.28}
\definecolor{richmaroon}{rgb}{0.69, 0.19, 0.38}
\definecolor{midnightblue}{rgb}{0.1, 0.1, 0.44}
\definecolor{rosebonbon}{rgb}{0.98, 0.26, 0.62}
\pdfoutput=1
\title{AdS correction to the Faddeev-Kulish state: Migrating from the Flat Peninsula}
\author[]{Sarthak Duary}
\affiliation[]{International Centre for Theoretical Sciences-TIFR,
	Shivakote, Hesaraghatta Hobli, Bengaluru North 560089, India}
\emailAdd{sarthak.duary@icts.res.in}
\date{}

\abstract{ The IR finiteness of $\mathcal{S}$-matrix in flat spacetime is tied to the Faddeev-Kulish dressed state, which suggests dressing the Fock space scattering state with the soft photon modes. We instigate a construction for the Faddeev-Kulish dressed state in the language of AdS/CFT. A salient feature of AdS spacetime is that it acquits itself as a quintessential IR regulator. The IR divergences will take shape after taking the zoomed in limit. We explore the Faddeev-Kulish dressed state to account for the AdS radius corrections. The Wilson line dressing stands as a guiding principle in the study of AdS radius-corrected Faddeev-Kulish dressing. We construct the modes of the Wilson line dressed massive scalar field implementing \textit{``vanilla HKLL reconstruction''} since the field is simply free field. This simplification is owing to the use of soft photon modes in the Wilson line dressing. We map the AdS radius-corrected soft photon modes in terms of CFT current operators. We invert this mapping, use the mapping in the Wilson line dressing, and express the AdS radius-corrected Faddeev-Kulish dressed state.}

\begin{document}
\maketitle
	
	\section{Introduction}
	Scattering amplitudes in QED vanish in four spacetime dimensions in flat space because of IR divergences. The soft photon interchange between the external legs due to long-range interactions is what causes these IR divergences. In perturbation theory, to all loop orders, the scattering process suffers from IR divergences. The soft contribution of each of the diagrams exponentiates after resumming the series\cite{Weinberg:1965nx}. Therefore, the non-perturbative amplitude, $\mathcal{A}$ vanishes after taking the infrared regulator, $\Lambda^{\text{reg}}_{\text{IR}}$ to zero
	\begin{equation}
	\mathcal{A} \to 0~~,~~~\Lambda^{\text{reg}}_{\text{IR}} \to 0~~~.
	\end{equation} 
	This means the probability of scattering is essentially nil. This is a quantum mechanical statement but this is really just a reflection of a classical fact that the power or energy radiated does not vanish for soft photons, basically soft photons are produced in order to match the classical answer \cite{Mott, Bloch}. The typical textbook solution to this IR divergence problem is to consider inclusive cross sections \cite{Bloch, Donoghue:1999qh} by taking a trace over the soft modes of the photons in the scattering states. This trace shifts the zero and yields a finite quantity. Since the trace is determined by the detector resolution, this is typically fine for phenomenological scenarios. Numerous soft photon modes evade detection and are thus regarded as unobservable. Nevertheless, in order to explore fine-grained issues regarding the unitarity of $\mathcal{S}$-matrix while taking soft modes of photon into consideration, $\mathcal{S}$-matrix must be defined appropriately\cite{Carney:2018ygh}.
	
	An upshot of the resolution of the IR divergence alternative to employing inclusive cross sections is to use ``dressed states'' as physical scattering states. While computing the $\mathcal{S}$-matrix, the ``in'' and ``out'' scattering states we choose reside in a Fock space. This choice of approximation is a nice one since while looking at timelike infinity (for massive particles) or at null infinity (for massless particles), particles are so far apart from one other that they barely interact and therefore are free. Nevertheless, in this scenario, the states corresponding to the Fock space basis is the ``sick basis'', leading the $\mathcal{S}$-matrix to become IR divergent. In order to construct an IR finite $\mathcal{S}$-matrix, the basis of scattering states has to be modified to incorporate the soft modes of the photons. The intuition is that because electro-magnetic interactions are long-range interactions, soft modes of photons in the ``in'' and ``out'' scattering states are always present. These dressed states by soft modes of the photons is referred as Faddeev-Kulish dressed state \cite{Kulish:1970ut}. The Faddeev-Kulish state is such that it precisely cancels the IR divergences in the $\mathcal{S}$-matrix, resulting in an IR finite $\mathcal{S}$-matrix \cite{Chung:1965zza, Kibble:1968oug, Kibble:1968npb, Kibble:1968lka, Ware:2013zja}. Recently, it was observed that the Faddeev-Kulish state arises as a consequence of asymptotic symmetries \cite{Strominger:2013jfa, Kapec:2015ena, Campiglia:2015qka, Campiglia:2015kxa, Hawking:2016sgy}, which indicates the existence of selection sectors \cite{Gabai:2016kuf, Kapec:2017tkm, Choi:2017bna, Choi:2017ylo}. For discussion of Faddeev-Kulish dressing in celestial space which is achieved with dressing by edge modes, see ref.\cite{Kapec:2021eug, Arkani-Hamed:2020gyp, PipolodeGioia:2022exe}. For a more recent discussion of Faddeev-Kulish state, see ref. \cite{Prabhu:2022zcr}.  
	
	In this paper, our goal is to explore the AdS radius correction to the Faddeev-Kulish state. We construct the Faddeev-Kulish dressing in AdS/CFT from the standpoint of the Wilson line dressing. The equivalence of the Faddeev-Kulish dressing and the Wilson line dressing involving the soft modes of photons or gravitons has been studied in \cite{Mandelstam:1962mi, Jakob:1990zi, Choi:2018oel, Choi:2019fuq}.\footnote{In flat spacetime, the Wilson line path is time-like geodesic for massive scattering states and for this geodesic, the Wilson line dressing describes the Faddeev-Kulish dressing\cite{Mandelstam:1962mi, Jakob:1990zi, Choi:2018oel, Choi:2019fuq}.}           
	
	We now talk about observables connected to the scattering process in the AdS/CFT and how it relates to observables in flat spacetime. The boundary observables in the AdS, thanks to the AdS/CFT correspondence are CFT correlation functions. The CFT correlation functions can be computed using the Witten diagrams in the bulk of AdS spacetime. Due to the absence of boundaries, the flat spacetime observables is defined asymptotically. The particular observable for scattering amplitude in flat spacetime is the $\mathcal{S}$-matrix. Zooming in around the center of the AdS spacetime, AdS spacetime manifests itself into flat spacetime. In this zommed in limit, the $\mathcal{S}$-matrix can be obtained from the CFT correlation functions using (i) Position space, (ii) Mellin space, and (iii) Momentum space representations of the CFT correlation functions, see ref. \cite{Polchinski:1999ry, Susskind:1998vk, Giddings:1999qu, Giddings:1999jq, Gary:2009mi, Penedones:2010ue, Fitzpatrick:2011jn, Fitzpatrick:2010zm, Gary:2009ae} for earlier developments, see ref.\cite{Komatsu:2020sag, vanRees:2022itk, Raju:2012zr, Li:2021snj, Okuda:2010ym, Maldacena:2015iua, Chandorkar:2021viw, Gadde:2022ghy} for recent developments.
Different CFT representations like Position space  \cite{Okuda:2010ym,Maldacena:2015iua,Komatsu:2020sag, vanRees:2022itk}, Mellin space \cite{Penedones:2010ue,Fitzpatrick:2011hu,Paulos:2016fap}, and Momentum space \cite{Raju:2012zr, Gadde:2022ghy}\ elucidate different roadmap to reach to the flat space $\mathcal{S}$-matrix. In recent years, the flat limit of Position space CFT correlation function using HKLL bulk reconstruction has been studied in \cite{Hijano:2019qmi, Hijano:2020szl} to realize IR sector physics in flat spacetime from techniques in AdS/CFT.
	This is essentially a statement of deciphering the physics of flat spacetime that is already stored in AdS/CFT.  In AdS spacetime, causality is attributed to analyticity and unitarity of CFT correlation functions \cite{Hofman:2008ar, Kelly:2014mra, Hartman:2015lfa, Hartman:2016dxc, Hofman:2016awc, Faulkner:2016mzt, 2017JHEP...07..066H, Afkhami-Jeddi:2016ntf, Belin:2019mnx, Caron-Huot:2020adz }. These conditions in the flat space $\mathcal{S}$-matrix have spawned intriguing implementations in the ``$\mathcal{S}$-matrix Bootstrap program'' from tools in ``CFT Bootstrap program'' \cite{Fitzpatrick:2011hu, Fitzpatrick:2011dm, Paulos:2016fap, Paulos:2016but, Paulos:2017fhb, Homrich:2019cbt, Gillioz:2020mdd, Kruczenski:2022lot}. In a very recent paper \cite{Hartman:2022njz}, focusing bound from ANEC for the CFT correlation function and it's connection to the flat space $\mathcal{S}$-matrix is studied.  
	
	Now, we turn to the discussion on Faddeev-Kulish dressing. Recently, we study the Faddeev-Kulish dressing in the earlier paper \cite{Duary:2022pyv} from the zoomed in limit of AdS/CFT. In this paper, we explore the AdS radius corrections to the Faddeev-Kulish dressed state which captures the outcome of cosmological constant on the flat spacetime state. We know the fact of life that the AdS radius acts as an IR regulator. In the zoomed in limit, the scattering amplitudes will have IR divergences. If we consider the scattering states dressed by the soft modes of the photons (Faddeev-Kulish dressed state), then we can get rid of the IR divergences. With these dressed states, we need to understand how the $\mathcal{S}$-matrix becomes the IR finite one from AdS/CFT. To understand this, first we need to understand how IR divergences manifest themselves after taking the zoomed in limit from the CFT correlation function using the Fock-space scattering states. The AdS radius corrections to this Faddeev-Kulish dressed state will provide new insight into an IR finite $\mathcal{S}$-matrix. 
	
	Now, we discuss the strategy to construct the AdS radius correction to the Faddeev-Kulish dressed state. As discussed before, we choose the Wilson line dressing as our guiding principle to arrive at the Faddeev-Kulish dressing in AdS/CFT. We are interested in the kinematic regime in which the scalar field is dressed by the soft modes of the photons. The dressed scalar field are free fields and the modes of the field can be reconstructed simply implementing the vanilla HKLL reconstruction. Upon taking the zoomed in limit, the creation/annihilation modes of the dressed field can be expressed in terms of the CFT operator corresponding to the undressed field which is dressed by the boundary-to-boundary Wilson line, which is the CFT representation. We can study the flat space representation taking into account AdS radius corrections as well by reexpressing the CFT operator corresponding to the undressed field in terms of the undressed mode of the scalar field. In order to have a fully fledged flat space representation, we must also express the Wilson line operator having CFT current operator in terms of photon creation/annihilation modes. In this paper, we explore AdS radius correction to the Faddeev-Kulish dressed state. To accomplish so, we have to invert the map between the soft modes of the photon as a smearing of the CFT current operators which shows up in the Wilson line operator. The $1/L^2$ corrected modes of the photon is derived in the paper \cite{Banerjee:2022oll}. The annihilation operator of an outgoing photon of negative helicity is given by   
	\begin{equation}
	\begin{split}
	&\sqrt{2\omega_{\vec{q}}}~\mathbf a^{\text{AdS} (-)}_{\vec{q}}\\
	&=\frac{1}{32\pi\omega_{\vec{q}}^2L^2 }\frac{1+z_q\bar{z}_q}{\sqrt{2}\omega_{\vec{q}}}\int d\tau^{\prime}~e^{-i\omega_{\vec{q}}L\left(\frac{\pi}{2}-\tau^{\prime}\right)}\int d^2z^{\prime} \int d^2z_w
	\Bigg[\frac{(1+z^{\prime}\bar{z}^{\prime})^2(1+z_w\bar{z}_w)^2}{(\bar{z}_w-\bar{z}^{\prime})^2(z_q-z_w)^3}\Bigg]\\
	&~~~~~~~~\partial_{z^{\prime}}j^-_{\bar{z}^{\prime}}(\tau^{\prime},z^{\prime},\bar{z}^{\prime})~,
	\end{split}
	\end{equation}  
	which is expressed in terms of a CFT current operator smeared over the boundary $S^2$. The $1/L^2$ corrected mode denoted by $\mathbf a_{\vec{q}}^{\text{AdS}(-)}$ is physically distinct from that of flat space mode of the photon. Here, there is an extra $z_w$ integral correponding to integration over intermediate angles $(z_w, \bar{z}_w)$ which is absent in the flat space mode. This result is used to study properties of the $1/L^2$ corrected $\mathcal{S}$-matrix from CFT physics, like soft photon theorem is derived from a CFT Ward identity \cite{Banerjee:2022oll}, which takes into consideration scattering in an asymptotically $AdS_4$ spacetimes where the $\mathcal{S}$-matrix is defined in the zoomed in region where the flat spacetime is located. 
	
	Now, in this paper we use these AdS corrected modes to express the Faddeev-Kulish dressed state. The creation mode of the soft Wilson line dressed massive scalar field is constructed implementing vanilla HKLL reconstruction since the dressed field is simply free field. The result we highlight in eq.\eqref{expr} is the expression for the dressed creation mode expressed in terms of the undressed creation mode with smearing over frequency and global time coordinate. The dressed mode dressed by the soft modes of photon acting on the vacuum state $\ket{0}$ is the Faddeev-Kulish state. The expression is the following 
	\begin{equation}
	\label{expr} 
	\begin{split}
	\sqrt{2\omega_{\vec{p}}}~\widetilde{a}^{\dagger}_{\omega_{\vec{p}}}
	&=\widetilde{\mathfrak{C}}~L \int d\Delta_{\vec{p}} ~e^{-i \Delta_{\vec{p}}L\Big[ \frac{\pi}{2}+\frac{i}{2}\log\Big(\frac{\Delta_{\vec{p}}+m}{\Delta_{\vec{p}}-m}\Big)\Big]}~ e^{i\omega_{\vec{p}}L\Big[\frac{\pi}{2}+\frac{i}{2}\log\Big(\frac{\omega_{\vec{p}}+m}{\omega_{\vec{p}}-m}\Big)\Big]}\\
	&~~~~~~~~\int d\tau ~e^{-iL \tau(\omega_{\vec{p}}-\Delta_{\vec{p}})}~e^{iq \int_{\Gamma(\tau,\hat{p})}j_a dx^a}~\sqrt{2\Delta_{\vec{p}}} ~\frac{a^{\dagger}_{\Delta_{\vec{p}}}}{\mathcal{C}}~~~.
	\end{split}
	\end{equation}

\begin{figure}[H]
\centering
\includegraphics[width=1\linewidth]{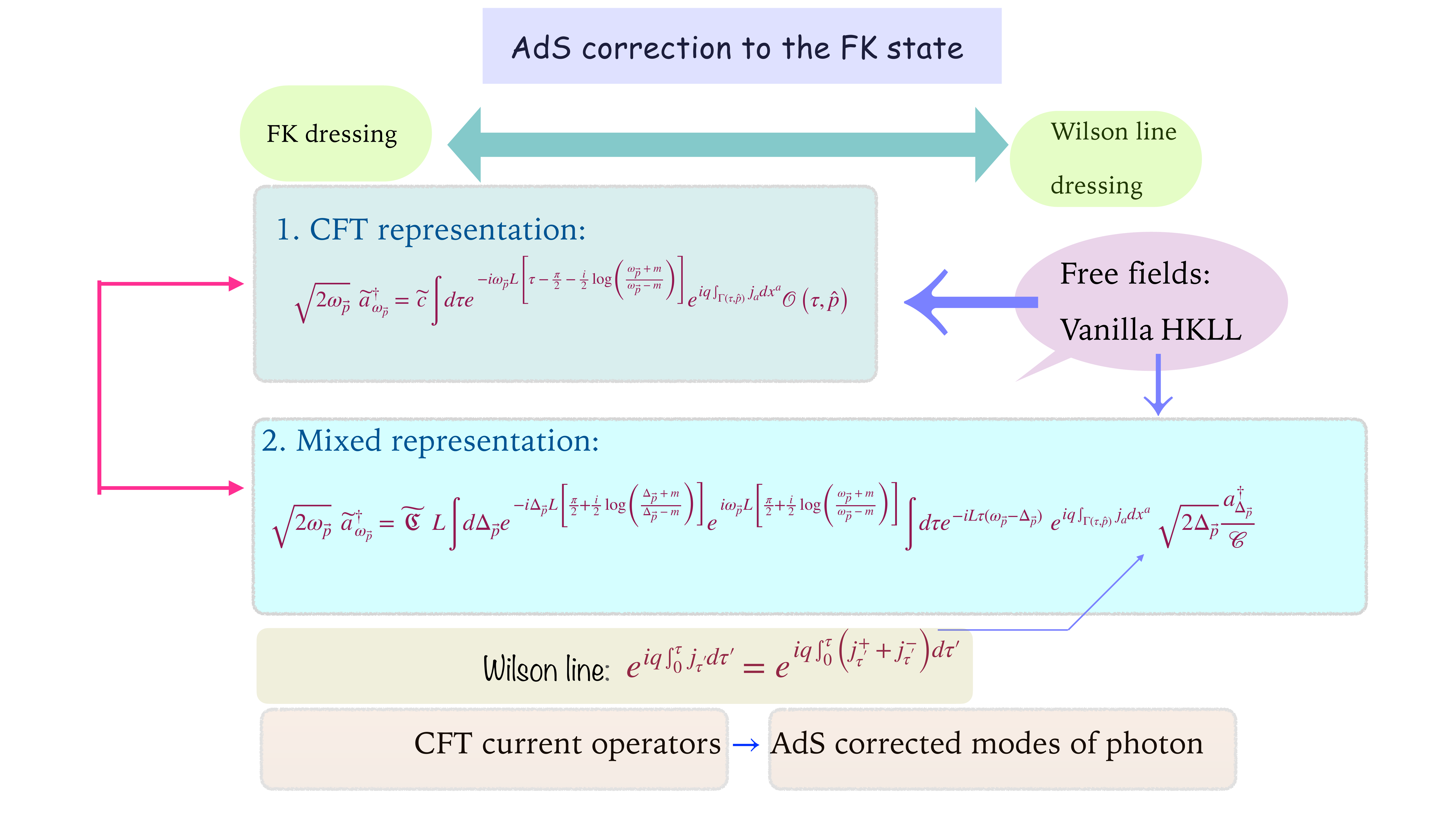}
\caption{Flowchart: Roadmap of the AdS correction to the Faddeev-Kulish(FK) state}
\label{fig:flowchart}
\end{figure}

	Now, this expression of eq.\eqref{expr} is written in terms of creation mode of the scalar field but the Wilson line which is expressed in terms of the CFT current operator. Therefore, this is in the ``mixed representation'' (mixed between the ``CFT representation'' and the ``flat space representation''). To write this dressed creation mode in the full fledged flat space representation, we have to express the CFT current operators in the Wilson line in terms of creation/annihilation operators of photon. We express the Wilson line for a particular path where global time coordinate varies from $0$ to $\tau$ keeping the angular direction fixed
	\begin{equation}
	\begin{split}
	e^{iq \int_{\Gamma(\tau,\hat{p})}j_a dx^a}=e^{iq\int_0^{\tau}\left(j^+_{\tau^{\prime}}+j^-_{\tau^{\prime}}\right)d\tau^{\prime}}~~~,
	\end{split}
	\end{equation}
	where, $j^+_{\tau^{\prime}}$ and $j^-_{\tau^{\prime}}$ corresponds to the positive and negative frequency modes of the photon when mapped to the photon modes. We evaluate the CFT current operators in terms of AdS corrected photon creation/annihilation modes. After that, we find the inverse mapping. That means we express the CFT current operators in terms of the AdS corrected modes of the photon. Now, choosing a particular path we  evaluate the Wilson line and for that we express the  global time component of the CFT current operator in terms of the photon creation/annihilation modes. Finally, we express the dressed creation mode of the scalar field in terms of the AdS corrected creation/annihilation modes of the photon. The dressed creation operator acting on the vacuum state gives the AdS radius-corrected Faddeev-Kulish dressed state. The roadmap is summerized in the flowchart fig.\ref{fig:flowchart}.

	\paragraph{\bf Organization of the paper.}
	The paper is organized as follows. In the section \ref{prelim} we review various things to make the paper self-contained. First, in the section \ref{flatlimit}, we discuss the zoomed in limit. We discuss how to extract the creation and annihilation operators for the free massive scalar field in terms of the CFT operators in this zoomed in limit. We use the equivalence between the Faddeev-Kulish dressing and the Wilson line dressing as our main strategy to construct the AdS correction to the Faddeev-Kulish dressed state. In the section \ref{Soft}, we study the soft Wilson line dressed field in AdS and evaluate the CFT operator corresponding to this dressed field. The soft Wilson line dressed scalar field turns itself into the free field and thus can be reconstructed implementing the vanilla HKLL reconstruction. In the section \ref{CFTrep}, we study the CFT representation as well as the mixed representation of the Faddeev-Kulish dressed state. In the section \ref{spec}, we talk about some speculations to get an IR finite $\mathcal{S}$-matrix from AdS/CFT. In the section \ref{adscorr}, we calculate the AdS corrected Faddeev-Kulish dressed state. To do this, first we express the CFT current operators in terms of AdS radius-corrected photon creation/annihilation operators in the section \ref{ADSC}. Next, in the section \ref{INV}, we evaluate the inverse mapping,  which implies we express the CFT current operators in terms of the AdS radius-corrected modes of the photon. Next, in the section \ref{tau}, we express the global time component of the CFT current operator in terms of the photon creation/annihilation modes. The global time component is needed since we choose a particular path to evaluate the Wilson line. Then, We express the dressed creation operator in terms of the AdS radius-corrected creation/annihilation modes of the photon. The dressed creation operator acting on the vacuum gives the AdS radius-corrected Faddeev-Kulish dressed state. Finally, we save the section \ref{Conclusions} for summarizing our conclusions and discussing future directions.  
	
	\section{Preliminaries}
	\label{prelim}
	In this section, we revisit some known things in order to make the paper self-contained. 
	\subsection{Flat Peninsula inside AdS Lake: The zoomed in limit}
	\label{flatlimit}
	In this section, we review the zoomed in limit of AdS/CFT and how to extract the creation and annihilation operators for the massive scalar scalar field in terms of the CFT operator in this limit using the vanilla HKLL bulk reconstruction described in \cite{Hijano:2020szl}. 
	
	At the level of geometry, we take the large AdS radius limit such that the global $AdS_4$ metric becomes that of flat spacetime. The $AdS_4$ lorentzian metric in global coordinates is given by 
	\begin{equation}
	ds^2=\frac{L^2}{\cos^2\rho} (-d\tau^2+d\rho^2+\sin^2\rho d\Omega_{2}^2)~~~.
	\end{equation}
	The coordinate replacement 
	\begin{equation}
	\tau=\frac{t}{L}~~\text{and}~~\rho=\frac{r}{L}~~~,\nonumber
	\end{equation}
	develops the $AdS_4$ metric into that of flat spacetime metric upon taking the limit $L \to \infty$
	\begin{equation}
	ds^2 \xrightarrow[L\to\infty]~-dt^2+dr^2+r^2d\Omega_{2}^2~~~.
	\end{equation} 
Essentially, the flat peninsula can be thought of as being inside the AdS lake, and this operation can be thought of as zoomed in limit.
	\subsection*{Massive scalar field modes in the zoomed in limit of AdS/CFT}
	Having discussed the zoomed in limit at the level of the geometry, we now turn to a discussion on fields. The creation and annihilation modes for a free massive scalar field in flat spacetime can be constructed in terms of the CFT operator in the zoomed in limit of AdS/CFT. 
	For normalizable modes of the bulk AdS scalar field, the bulk AdS scalar field $\phi(\rho,x)$ is related to the dual boundary CFT operator $\mathcal{O}(x)$ through the fall-off condition  
	\begin{equation}
	\phi(\rho,x) \xrightarrow[\rho\to\frac{\pi}{2}]~(\cos\rho)^{\Delta} \mathcal{O}(x)~~~.
	\end{equation} 
	We have 
	\begin{equation}
	\begin{split}
	m^2L^2&=\Delta(\Delta-3)\\
	\implies \Delta&=\frac{3}{2}+mL+\mathcal{O}(L)^{-1}~~~.
	\end{split}
	\end{equation}
	To extract the creation and annihilation modes in flat spacetime in terms of CFT operators, first we reconstruct bulk operators as operators in the CFT using the vanilla HKLL bulk reconstruction prescription. The AdS scalar operator has been sent to the scattering region of the flat spacetime  
	\begin{equation}
	\tau=\frac{t}{L}~~~\text{and} ~~~\rho=\frac{r}{L}~~~. 
	\end{equation}
	Free scalar fields in flat spacetime can be expanded in creation/annhilation modes. The scalar field can be mode expanded as  
	\begin{equation}
	\phi(x)=\int \frac{d^3\vec{p}}{(2\pi)^3}\frac{1}{\sqrt{2\omega_{\vec{p}}}} (a^{}_{\omega_{\vec{p}}}e^{ip.x}+a^{\dagger}_{\omega_{\vec{p}}}e^{-ip.x})~~~.
	\end{equation} 
	We can extract the creation/annhilation modes of the scalar field from the position space field operator of the scalar field. The approach we follow is the following. First, we construct the free local bulk operators in the CFT using the free HKLL bulk reconstruction. Next, we extract the creation/annhilation modes using fourier transform and then we take a large AdS radius limit which is the zoomed in limit. The flat space creation mode for the free massive scalar field $\phi$ is given by \cite{Hijano:2020szl}
	\begin{equation}
	\label{massivecreation}
	\sqrt{2\omega_{\vec{p}}}~a^{\dagger}_{\omega_{\vec{p}}}=\mathcal{C} \int d\tau^{\prime} e^{-i \omega_{\vec{p}} L\left[\tau^{\prime}-\frac{\pi}{2}-\frac{i}{2}\log\Big(\frac{\omega_{\vec{p}}+m}{\omega_{\vec{p}}-m}\Big)\right]}\mathcal{O}\left(\tau^{\prime} ,\hat{p}\right)~~~,
	\end{equation}
	where 
	\begin{equation}
	\begin{split}
	\mathcal{C}&=\frac{1}{2\pi} \Bigg(\frac{mL}{\pi^3}\Bigg)^{\frac{1}{4}} \Bigg(\frac{2m}{i |\vec{p}|}\Bigg)^{mL+\frac{1}{2}}~ L~~~.\\
	\end{split}
	\end{equation} 
	In the formula of eq.\eqref{massivecreation}, the exponential part in the integrand is highly oscillatory as we take the zoomed in limit, $L \to \infty$, therefore the insertion points of the operators are in windows of size $\mathcal{O}(1/L)$ at the
	complex points 
	\begin{equation}
	\text{Re}(\tau^{\prime})=\frac{\pi}{2} ~~,~\text{and}~~~
	\text{Im}(\tau^{\prime})=\frac{1}{2}\text{log}\Bigg(\frac{\omega_{\vec{p}}+m}{\omega_{\vec{p}}-m}\Bigg) ~~~ .
	\end{equation}
	\subsection{Soft Wilson line dressed scalar field in AdS and dual CFT operator}
	\label{Soft}
	In this section, we review the set up of the Faddeev-Kulish dressed state in the zoomed in limit of AdS/CFT from the earlier paper \cite{Duary:2022pyv}. 
	
	First, we explain the soft Wilson line dressed scalar field in AdS and evaluate the CFT operator corresponding to this soft Wilson line dressed field.
	The bulk massive scalar field dressed by the bulk-to-boundary Wilson line $U_{\mathfrak{B}\partial}(y,x)$ is given by   
	\begin{equation}
	\widetilde\phi(y)=U_{\mathfrak{B}\partial}(y,x)\phi(y)~~~,
	\end{equation}
	where the bulk-to-boundary Wilson line is 
	\begin{equation}
	\begin{split}
	U_{\mathfrak{B}\partial}(y,x)&=\mathcal{P}\Bigg\{e^{iq\int_{\Gamma}^{} dx^M A_M}\Bigg\}\\
	&=\mathcal{P}\Bigg\{e^{iq\int_{y}^{x} dx^M A_M}\Bigg\}
	~~~.
	\end{split}
	\end{equation}
	Here, $\Gamma$ is the path in AdS that joins bulk point point $y$ to boundary point $x$, $y \to x$. Now, considering scalar electrodynamics with action 
	\begin{equation}
	S=\int d^4x \sqrt{-g}\Big(-D_M\phi^{\dagger}D^M\phi-\frac{1}{4}F_{MN}F^{MN}-m^2\phi^{\dagger}\phi\Big)~,
	\end{equation}
	where $D_M=\partial_M-iqA_M$, the equation of motion of the scalar field is given by 
	\begin{equation}
	\left(\square-m^2\right)\phi=iq(\phi\nabla_MA^M+2A^M\partial_M\phi)+q^2A^2\phi~~~.
	\end{equation} 
	Now, the Wilson line dressed field $\widetilde{\phi}$ satisfies \cite{Guica:2015zpf}   
	\begin{equation}
	\small 
	\label{dressedeom}
	\begin{split}
	\left(\square-m^2\right)\widetilde{\phi}=-iq\widetilde{\phi}\nabla^M\int_{\Gamma}F_{MP}dy^P-2iq\nabla^M\widetilde{\phi}\int_{\Gamma}F_{MP}~dy^P+q^2\widetilde{\phi}~g^{MN}\int_{\Gamma}F_{MP}dy^P\int_{\Gamma}F_{NQ}dy^Q~~~.
	\end{split}
	\end{equation}
	We choose the Wilson line dressing in such a way that the field strength $F_{MN}$ becomes $\mathcal{O}\left(\frac{1}{L}\right)$. We dress the scalar field with soft modes of the photon in the Wilson line and in AdS, the minimum frequency of photon is of  $\mathcal{O}\left(\frac{1}{L}\right)$. This dressed scalar field we refer as soft Wilson line dressed scalar field. As a consequence of this dressing, the field $\widetilde{\phi}$ is free field in the zoomed in limit.\footnote{There is an alternative way to get the Faddeev-Kulish dressed state by considering a `Soft-collinear effective theory(SCET)' lagrangian \cite{Becher:2014oda} to construct an asymptotic Hamiltonian. While constructing Faddeev-Kulish states we consider the soft part only, and put a hard cutoff on the photon energy. See ref. \cite{Hannesdottir:2019rqq} for the construction of Faddeev-Kulish states within the framework of SCET.}
	Using this simplification, $\widetilde{\phi}$ can ve obtained using vanilla HKLL. The boundary CFT operator dual to the Wilson line dressed field $\widetilde{\phi}$ we denote by $\widetilde{{\mathcal{O}}}$ and the operator $\widetilde{{\mathcal{O}}}$ is non-local since it involves boundary-to-boundary Wilson line. The Wilson line dressed scalar field $\widetilde{\phi}$ is given by 
	\begin{equation}
	\widetilde{\phi}(y)=\mathcal{P}\Bigg\{e^{iq\int_{y}^{x}A_M dx^M}\Bigg\}\phi(y)~~~.
	\end{equation}
	The fall-off conditions of the photon field and the scalar bulk fields are 
	\begin{equation}
	\begin{split}
	&A_a\left(\rho,x\right)~\xrightarrow[\rho\to\frac{\pi}{2}]~j_a(x)~\cos\rho~~~\\
	&\phi\left(\rho,x\right)~\xrightarrow[\rho\to\frac{\pi}{2}]~(\cos\rho)^{\Delta}\mathcal{O}(x)~~~\\
	&\widetilde{\phi}\left(\rho,x\right)~\xrightarrow[\rho\to\frac{\pi}{2}]~(\cos\rho)^{\Delta}\mathcal{\widetilde{O}}(x)~~~.
	\end{split}
	\end{equation}
	Therefore, the CFT operator at the boundary corresponding to the soft Wilson line dressed field is given by 
	\begin{equation}
	\label{bnd}
	\widetilde{\mathcal{O}}(x^{\prime})=U_{\partial \partial}(x^{\prime},x) \mathcal{O}(x^{\prime})~~~.
	\end{equation}
	The boundary-to-boundary Wilson line $U_{\partial \partial}$ is given by
	\begin{equation}
	U_{\partial \partial}(x^{\prime},x) =\mathcal{P}\Bigg\{e^{i q \int_{x^{\prime}}^{x} dx^a j_a}\Bigg\}~~~.
	\end{equation} 
	The eq.\eqref{bnd} can be straightway derived using the fall-off behaviors of the fields. To derive this equation, there is a cute cancellation between the fall-off of the photon field and the line element factor $(\cos \rho)^{-1}$ going from bulk to boundary line element. 
	\section{CFT and Mixed representations of the Faddeev-Kulish dressed state: Vanilla HKLL}
	\label{CFTrep}
	\subsection{CFT representation}
	\label{cft}
	In this section, we express the Faddeev-Kulish dressed state in CFT representation. 
	As discussed in the previous section, considering the field strength $F_{MN}= \mathcal{O}\left(\frac{1}{L}\right)$, the Wilson line dressed scalar field $\widetilde{\phi}$ is free field in the zoomed in limit and can be obtained using vanilla HKLL. The creation mode for the free massive scalar field $\widetilde{\phi}$ was constructed in \cite{Hijano:2020szl} and it is given by
	\begin{equation}
	\label{creationmode}
	\sqrt{2\omega_{\vec{p}}}~\widetilde{a}^{\dagger}_{\omega_{\vec{p}}}=\widetilde{c} \int d\tau e^{-i \omega_{\vec{p}} L\left[\tau-\frac{\pi}{2}-\frac{i}{2}\log\Big(\frac{\omega_{\vec{p}}+m}{\omega_{\vec{p}}-m}\Big)\right]}\widetilde{\mathcal{O}}\left(\tau,\hat{p}\right)~~~.
	\end{equation}
	where 
	\begin{equation}
	\begin{split}
	\widetilde{c}&=\frac{1}{2\pi} \Bigg(\frac{mL}{\pi^3}\Bigg)^{\frac{1}{4}} \Bigg(\frac{2m}{i |\vec{p}|}\Bigg)^{mL+\frac{1}{2}}~ L~~~.\\
	\end{split}
	\end{equation}
	Now, the CFT operator of the dressed scalar field $\widetilde{\mathcal{O}}\left(\tau,\hat{p}\right)$ is related to the CFT operator $\mathcal{O}\left(\tau,\hat{p}\right)$ as 
	\begin{equation}
	\label{tildeO}
	\widetilde{\mathcal{O}}\left(\tau,\hat{p}\right)
	=e^{iq \int_{\Gamma(\tau,\hat{p})}j_a dx^a}\mathcal{O}\left(\tau,\hat{p}\right)~~~.
	\end{equation}
	Using eq.\eqref{tildeO}, we get the creation mode of the dressed massive scalar field in terms of $\mathcal{O}\left(\tau,\hat{p}\right)$
	\begin{equation}
	\sqrt{2\omega_{\vec{p}}}~\widetilde{a}^{\dagger}_{\omega_{\vec{p}}}=\widetilde{c} \int d\tau e^{-i \omega_{\vec{p}} L\left[\tau-\frac{\pi}{2}-\frac{i}{2}\log\Big(\frac{\omega_{\vec{p}}+m}{\omega_{\vec{p}}-m}\Big)\right]} e^{iq \int_{\Gamma(\tau,\hat{p})}j_a dx^a}\mathcal{O}\left(\tau,\hat{p}\right)~~~.\\
	\end{equation} 
	The Wilson line dressed creation mode $\widetilde{a}^{\dagger}_{\omega_{\vec{p}}}$ acting on the CFT vacuum $\ket{0}$ is the CFT representation of the Faddeev-Kulish dressed state. 
	
\begin{figure}[H]	
		\centering
		\includegraphics[width=0.5\linewidth]{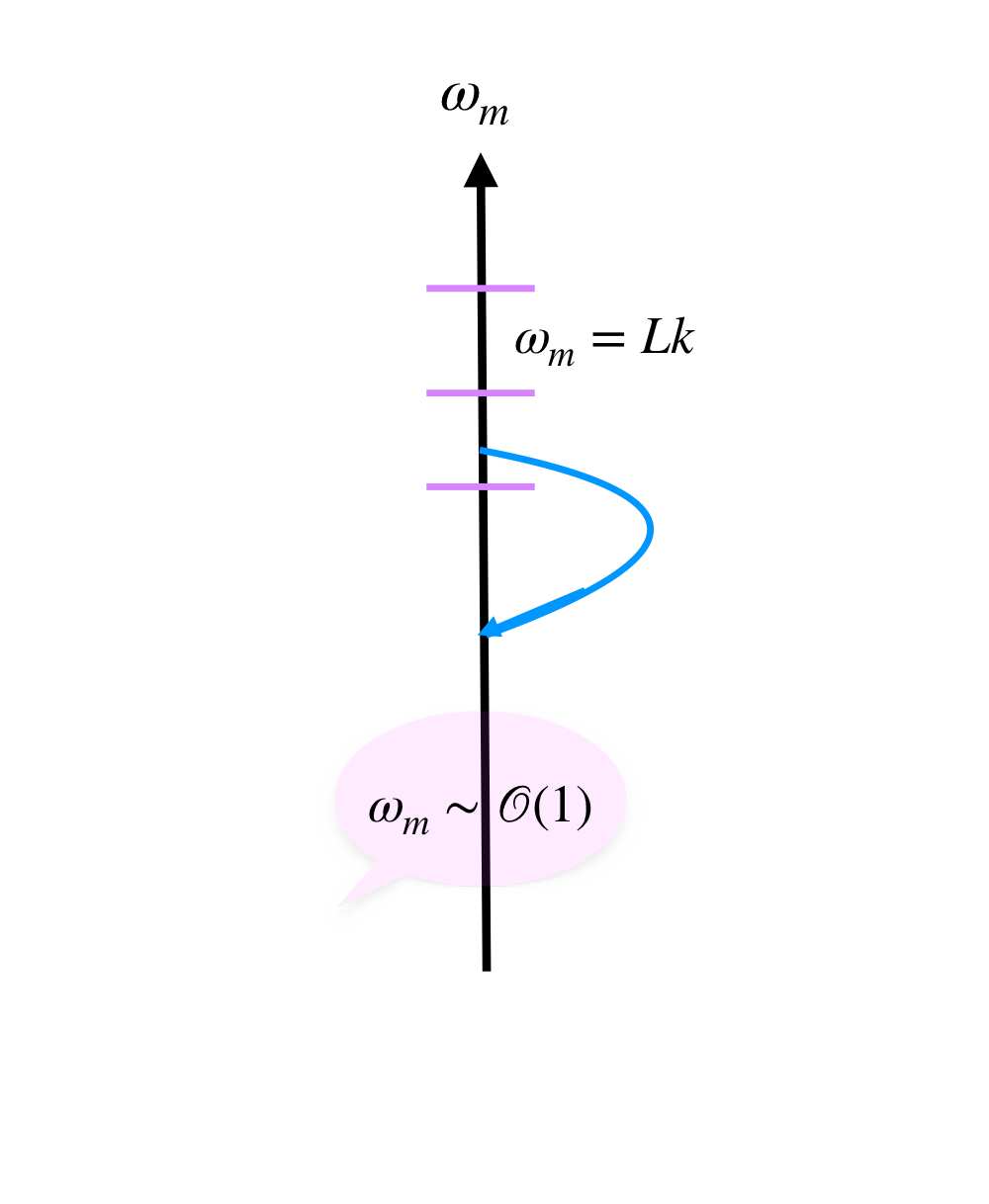}
		\caption{Frequency scale $\omega_m$: $\omega_m \sim \mathcal{O}(1)$ and $\omega_m=Lk$.}
		\label{fig:freqscale}
	\end{figure}

	Now, in the expression of the CFT current operator in the Wilson line we implement $F_{MN} \to 0$ limit and the Wilson line does not depend on path as a consequence of this limit. Therefore, this enables us to choose a particular path for the Wilson line $\Gamma(\tau,\hat{p})$ which connectes the two points in the boundary to evaluate $e^{iq \int_{\Gamma(\tau,\hat{p})}j_a dx^a}$. We choose the path to be global time coordinate varies from $0$ to $\tau$ while keeping the coordinates of the angular direction fixed. Now, we evaluate the global time integral. Denoting the global time component of the CFT current operator as
	\begin{equation}
	j_{\tau}(\tau^{\prime},\hat{p})=\sum_{m}^{}e^{i \omega_m \tau^{\prime}} \hat{\chi}_{m}(\hat{p})~~~,
	\end{equation}
	where $\hat{\chi}_{m}$ involves spherical harmonics $Y_l^m(\hat{\Omega})$. Here, we denote the sum over $\text{modes}$ by $m$. Integrating over global time yields
	\begin{equation}
	\int_{0}^{\tau}j_{\tau}(\tau^{\prime},\hat{p}) d\tau^{\prime}=\sum_{m}^{}\frac{1}{i\omega_m}  (e^{i\omega_m \tau}-1)\hat{\chi}_{m}(\hat{p})~~~.
	\end{equation}
	We expand the Wilson line operator $e^{iq \int_{0}^{\tau} j_{\tau}(\tau^{\prime},~\hat{p})~ d\tau^{\prime}}$ in series to $\mathcal{O}(q)$
	\begin{equation}
	\label{expand}
	\begin{split}
	e^{iq \int_{0}^{\tau} j_{\tau}(\tau^{\prime},~\hat{p})~ d\tau^{\prime}}
	&=1+q \sum_{m}^{}\frac{1}{\omega_m}  (e^{i\omega_m \tau}-1)\hat{\chi}_{m}(\hat{p})+\mathcal{O}(q^2)~~~,
	\end{split}
	\end{equation}
where, we use the expansion 
\begin{equation}
\begin{split}
e^{iq \int_{0}^{\tau} j_{\tau}(\tau^{\prime},~\hat{p})~ d\tau^{\prime}}&=1+iq\int_{0}^{\tau}j_{\tau}(\tau^{\prime},\hat{p}) d\tau^{\prime} +\mathcal{O}(q^2)~~~.
\end{split}
\end{equation}	
	Scalar/vector soft modes have frequency $\omega_m=\begin{Bmatrix}
	1 \\ 
	2 
	\end{Bmatrix}+l+2\kappa$, $\kappa\in \mathbb{Z}^{+}$.\footnote{We refer the reader to the paper \cite{Hijano:2020szl} (the section ``Reconstruction of $U(1)$ gauge fields in global $AdS_4$'') for details regarding the scalar/vector modes.} 
	Now, there are two different alternatives of the soft modes in the zoomed in limit. One is that $\kappa \sim {\cal O}(1)$, and the frequency $\omega_m/L \rightarrow 0$. The mode is given by
	\begin{equation}
	\omega_m \sim \mathcal{O}(1)~~~,~~~\frac{\omega_m}{L} \to 0~~~.
	\end{equation}
	Another alternative is to denote $\omega_m=kL$ and take the zoomed in limit, $L\to \infty$, and then to take soft limit with $k\rightarrow 0$. The mode is given by 
	\begin{equation}
	\omega_m=Lk~~~,~~~\frac{\omega_m}{L} = k \to 0~~~.
	\end{equation}
	In fig.\ref{fig:freqscale}, we draw a frequency scale to denote two different alternatives of the soft modes in the zoomed in limit. 
In the zoomed in limit, $\omega_m \sim 2\kappa=L k$ dominates the sum over modes, and therefore
\begin{equation}
\sum_{\kappa} \to \frac{L}{2} \int dk~~~.
\end{equation}	
	The Wilson line dressed creation mode of the massive scalar field at $\mathcal{O}(q)$ is given by 
	\begin{equation}
	\begin{split}
	\sqrt{2\omega_{\vec{p}}}~\widetilde{a}^{\dagger}_{\omega_{\vec{p}}}=\frac{q}{2}\widetilde{c} \int d\tau \int \frac{dk}{k}e^{-i\omega_{\vec{p}} L\left[\tau-\frac{\pi}{2}-\frac{i}{2}\log\left(\frac{\omega_{\vec{p}}+m}{\omega_{\vec{p}}-m}\right)\right]}\left(e^{iLk\tau}-1\right)\hat{\chi}_k(\hat{p})~{\mathcal{O}}\left(\tau,\hat{p}\right)~~~.
	\end{split}
	\end{equation}
	This expression for the dressed creation mode acting on CFT vacuum $\ket{0}$ state gives the ``CFT representation'' of the Faddeev-Kulish dressed state. We expect the $\mathcal{S}$-matrix constructed with the dressed scattering state will be IR-finite. In the next section \ref{spec}, we discuss some speculations on this issue. 
	
	\subsubsection{Some speculation to get an IR finite $\mathcal{S}$-matrix from AdS/CFT}
	\label{spec}
	In this section, we discuss some speculations to get an IR finite $\mathcal{S}$-matrix from AdS/CFT.  
	The CFT operator at the boundary corresponding to the soft Wilson line dressed field is given by   
	\begin{equation}
	\widetilde{\mathcal{O}}(x^{\prime})=\mathcal{P}\Bigg\{e^{i q \int_{x^{\prime}}^{x} dx^a j_a}\Bigg\} \mathcal{O}(x^{\prime})~~~.
	\end{equation}
	
	The Wilson line doesn't depend on the path and only depends on the endpoints of the path is precisely when the field strength vanishes. When we dress the scalar field $\phi$ with soft modes of photon, then this is exactly what happens. The boundary-to-boundary Wilson line given by 
	\begin{equation}
	U_{\partial \partial}(x^{\prime},x) =\mathcal{P}\Bigg\{e^{i q \int_{x^{\prime}}^{x} dx^a j_a}\Bigg\}~~~,
	\end{equation}  
	is dependent only on the endpoints.
	
	Now, The dressed $4$-point amplitude in terms of the undressed $4$-point amplitude is expressed as 
	\begin{equation}
	\expval{\widetilde{O} \widetilde{\mathcal{O}} \widetilde{\mathcal{O}} \widetilde{\mathcal{O}}}=\sum_{p, q, r,s}^{} \frac{1}{p! q! r! s!} \expval{\mathfrak{J}^p~ \mathcal{O} ~\mathfrak{J}^q~ \mathcal{O}~ \mathfrak{J}^r~ \mathcal{O}~ \mathfrak{J}^s~\mathcal{O}} ~~~,
	\end{equation}
	where,  we use the shorthand notation 
	\begin{equation}
	\mathfrak{J}^p \equiv \Bigg[iq \int_{x^{\prime}}^{x} dx^{ a} j_{a} \Bigg]^p~~~.
	\end{equation} 
	The line integral in the Wilson line runs from the insertion point of the boundary CFT operator to another point at the boundary. Now, the $4$-point amplitude will involve OPE between the operator $\mathcal{O}$ with CFT current operator $j_a$ since the original CFT operator of the undressed field $\mathcal{O}(x^{\prime})$ will have the insertion point $x^{\prime}$. The Wilson line depends on the end points therefore there is a CFT current operator $j_a(x^{\prime})$ at the insertion point of the CFT operator $\mathcal{O}$, dual to the undressed field.\footnote{See, for instance the OPE shows up naturally for constructing IR-finite celestial amplitude. (Ref. \cite{Afkhami-Jeddi:2016ntf} section 3.3.3. ``IR-finite celestial amplitude''.)} We expect that the undressed state which is the typical Fock-space state will bring IR divergence which will be neatly cancelled by the piece of the dressing in the ``CFT representation'' of the Faddeev-Kulish dressed state.   
	
	\subsection{Mixed representation}
	\label{mixed}
\begin{figure}[H]	
		\centering
		\includegraphics[width=0.5\linewidth]{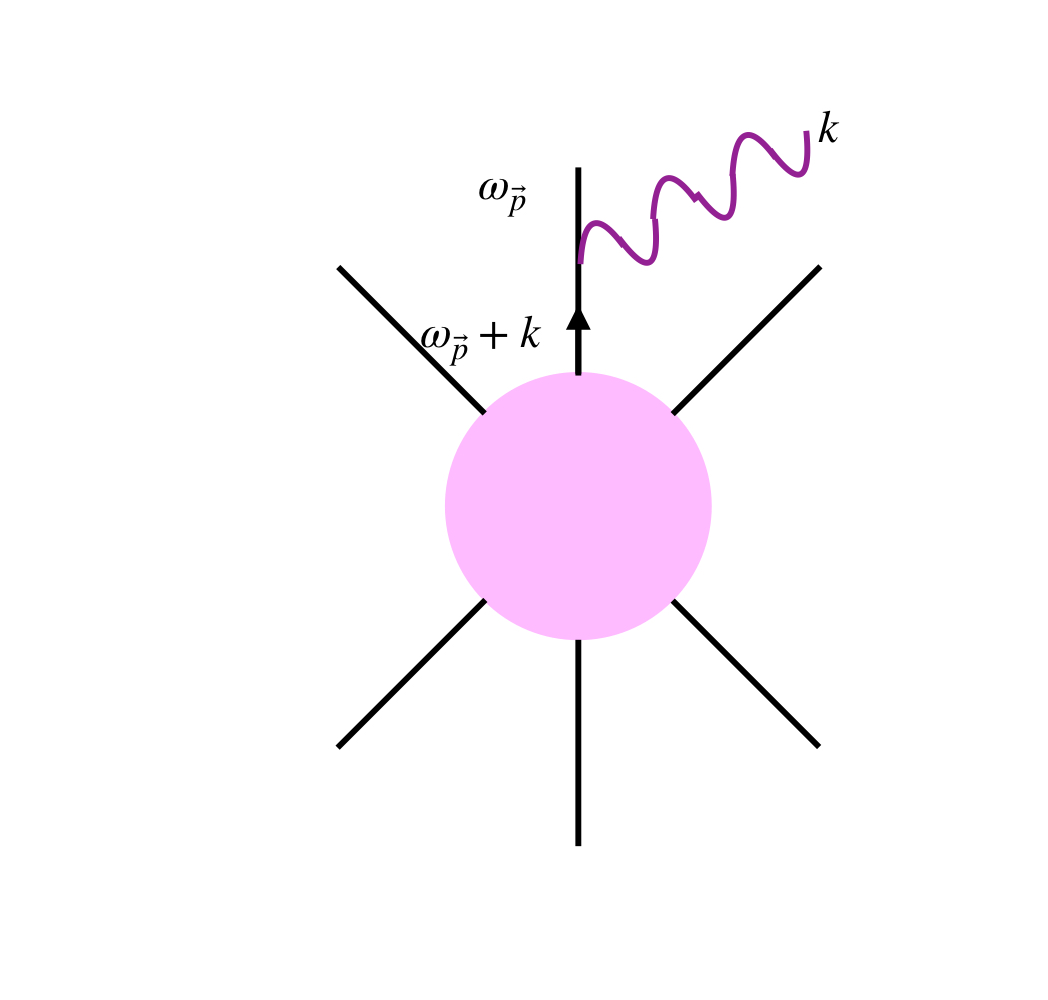}
		\caption{Dressing with soft modes of Wilson line: The frequency of the creation mode of the scalar field will get shifted from $\omega_{\vec{p}}$ to  $\omega_{\vec{p}}+k$ as a consequence of the dressing.}
		\label{fig:mixedrep}
	\end{figure}

	We express the creation mode for the Wilson line dressed massive scalar field with the soft modes of the photon in terms of undressed mode.
	To accomplish so, we insert a delta function $\delta(\tau-\tau^{\prime})$ and express the delta function in terms of integral over frequency, $\Delta_{\vec{p}}$ which is the corresponding frequency of the undressed mode of the massive scalar field  
	\begin{equation}
	\begin{split}
	\int d\tau^{\prime}  \delta(\tau^{\prime}-\tau)=\int d\tau^{\prime}\int d\Delta_{\vec{p}}~ \frac{L}{2\pi} e^{-i \Delta_{\vec{p}}L(\tau^{\prime}-\tau)}~~~.
	\end{split}
	\end{equation}
	Therefore, the creation mode of the soft Wilson line dressed massive scalar field is expressed as   
	\begin{equation}
	\label{creation}
	\begin{split}
	\sqrt{2\omega_{\vec{p}}}~\widetilde{a}^{\dagger}_{\omega_{\vec{p}}}
	&=\widetilde{c} \int d\tau e^{-i \omega_{\vec{p}} L\left[\tau-\frac{\pi}{2}-\frac{i}{2}\log\Big(\frac{\omega_{\vec{p}}+m}{\omega_{\vec{p}}-m}\Big)\right]} e^{iq \int_{\Gamma(\tau,\hat{p})}j_a dx^a}\mathcal{O}\left(\tau,\hat{p}\right)\\
	&=\widetilde{c} \int d\tau ~\Bigg[\int d\tau^{\prime} {\delta(\tau^{\prime}-\tau)}\Bigg]~e^{-i \omega_{\vec{p}} L\left[\tau-\frac{\pi}{2}-\frac{i}{2}\log\Big(\frac{\omega_{\vec{p}}+m}{\omega_{\vec{p}}-m}\Big)\right]} e^{iq \int_{\Gamma(\tau,\hat{p})}j_a dx^a}\mathcal{O}\left(\tau,\hat{p}\right)\\
	&=\widetilde{c} \int d\tau \int d\tau^{\prime} {\int d\Delta_{\vec{p}}~ \frac{L}{2\pi} e^{-i \Delta_{\vec{p}}L(\tau^{\prime}-\tau)}}~e^{-i \omega_{\vec{p}} L\left[\tau-\frac{\pi}{2}-\frac{i}{2}\log\Big(\frac{\omega_{\vec{p}}+m}{\omega_{\vec{p}}-m}\Big)\right]}\\ &~~~~~~~~~~e^{iq \int_{\Gamma(\tau,\hat{p})}j_a dx^a}\mathcal{O}\left(\tau^{\prime},\hat{p}\right)\\
	&=\widetilde{\mathfrak{C}}L \int d\Delta_{\vec{p}} ~e^{-i \Delta_{\vec{p}}L\Big[ \frac{\pi}{2}+\frac{i}{2}\log\Big(\frac{\Delta_{\vec{p}}+m}{\Delta_{\vec{p}}-m}\Big)\Big]}~\int d\tau e^{-i\omega_{\vec{p}}L\Big[\tau-\frac{\pi}{2}-\frac{i}{2}\log\Big(\frac{\omega_{\vec{p}}+m}{\omega_{\vec{p}}-m}\Big)\Big]} e^{i\Delta_{\vec{p}}L \tau}\\
	&~~~~~~e^{iq \int_{\Gamma(\tau,\hat{p})}j_a dx^a}\int d\tau^{\prime}~e^{-i \Delta_{\vec{p}}L\Big[ \tau^{\prime}-\frac{\pi}{2}-\frac{i}{2}\log\Big(\frac{\Delta_{\vec{p}}+m}{\Delta_{\vec{p}}-m}\Big)\Big]}~\mathcal{O}\left(\tau^{\prime},\hat{p}\right)~~~,
	\end{split}
	\end{equation}
	where, in the last step we use
	\begin{equation}
	e^{-i \Delta_{\vec{p}}L(\tau^{\prime}-\tau)}=e^{-i \Delta_{\vec{p}}L\Big[ \frac{\pi}{2}+\frac{i}{2}\log\Big(\frac{\Delta_{\vec{p}}+m}{\Delta_{\vec{p}}-m}\Big)\Big]}~e^{i\Delta_{\vec{p}}L \tau}~e^{-i \Delta_{\vec{p}}L\Big[ \tau^{\prime}-\frac{\pi}{2}-\frac{i}{2}\log\Big(\frac{\Delta_{\vec{p}}+m}{\Delta_{\vec{p}}-m}\Big)\Big]}~~~.
	\end{equation}
	where,
	\begin{equation}
	\begin{split}
	\widetilde{c}&=\frac{1}{2\pi} \Bigg(\frac{mL}{\pi^3}\Bigg)^{\frac{1}{4}} \Bigg(\frac{2m}{i |\vec{p}|}\Bigg)^{mL+\frac{1}{2}}~ L\\
	\widetilde{\mathfrak{C}}&=\frac{\widetilde{c}}{2 \pi}~~~.
	\end{split}
	\end{equation}
	Now, using the expression for the creation mode $a^{\dagger}_{\Delta_{\vec{p}}}$ corresponing to frequency of the undressed mode, $\Delta_{\vec{p}}$  in terms of the boundary CFT operator $\mathcal{O}\left(\tau^{\prime},\hat{p}\right)$   
	\begin{equation}
	\sqrt{2\Delta_{\vec{p}}}~a^{\dagger}_{\Delta_{\vec{p}}}=\mathcal{C}\int d\tau^{\prime}~e^{-i \Delta_{\vec{p}}L\Big[ \tau^{\prime}-\frac{\pi}{2}-\frac{i}{2}\log\Big(\frac{\Delta_{\vec{p}}+m}{\Delta_{\vec{p}}-m}\Big)\Big]}\mathcal{O}\left(\tau^{\prime},\hat{p}\right)~~~,
	\end{equation}
	eq.\eqref{creation} is expressed as 
	\begin{equation}
	\label{cr2}
	\begin{split}
	\sqrt{2\omega_{\vec{p}}}~\widetilde{a}^{\dagger}_{\omega_{\vec{p}}}
	&=\widetilde{\mathfrak{C}}~L \int d\Delta_{\vec{p}} ~e^{-i \Delta_{\vec{p}}L\Big[ \frac{\pi}{2}+\frac{i}{2}\log\Big(\frac{\Delta_{\vec{p}}+m}{\Delta_{\vec{p}}-m}\Big)\Big]}~ e^{i\omega_{\vec{p}}L\Big[\frac{\pi}{2}+\frac{i}{2}\log\Big(\frac{\omega_{\vec{p}}+m}{\omega_{\vec{p}}-m}\Big)\Big]}\\
	&~~~~~~~~\int d\tau ~e^{-iL \tau(\omega_{\vec{p}}-\Delta_{\vec{p}})}~e^{iq \int_{\Gamma(\tau,\hat{p})}j_a dx^a}~\sqrt{2\Delta_{\vec{p}}} ~\frac{a^{\dagger}_{\Delta_{\vec{p}}}}{\mathcal{C}}~~~.
	\end{split}
	\end{equation}
	In eq.\eqref{cr2}, we express the creation mode of the soft Wilson line dressed massive scalar field in terms of the undressed creation mode. We choose a particular path for the Wilson line $\Gamma(\tau,\hat{p})$ as in the previous section \ref{cft}. We mode expand the $\tau$ component of the CFT current operator as
	\begin{equation}
	j_{\tau}(\tau^{\prime},\hat{p})=\sum_{m}^{}e^{i \omega_m \tau^{\prime}} \hat{\chi}_{m}(\hat{p})~~~.
	\end{equation}
	Now, we perform the $\tau$ integral and get 
	\begin{equation}
	\begin{split}
	&\int d\tau ~e^{-iL \tau(\omega_{\vec{p}}-\Delta_{\vec{p}})}~e^{iq \int_{0}^{\tau} j_{\tau}(\tau^{\prime},~\hat{p})~ d\tau^{\prime}}\\
	&=\int d\tau ~e^{-iL \tau(\omega_{\vec{p}}-\Delta_{\vec{p}})}~\Bigg[1+q \sum_{m}^{}\frac{1}{\omega_m}  (e^{i\omega_m \tau}-1)\hat{\chi}_{m}(\hat{p})+\mathcal{O}(q^2)\Bigg]\\
	&=\frac{2\pi}{L} \delta(\omega_{\vec{p}}-\Delta_{\vec{p}})+\frac{2\pi}{L} \sum_{m}^{}\frac{~ q~\hat{\chi}_m}{\omega_m}\Bigg(\delta\Big[ (\omega_{\vec{p}}-\Delta_{\vec{p}})+\frac{\omega_m}{L}\Big]-\delta(\omega_{\vec{p}}-\Delta_{\vec{p}})\Bigg)+\mathcal{O}(q^2)~~~.
	\end{split}
	\end{equation}
	Scalar/Vector soft modes have frequency $\omega_m=\begin{Bmatrix}
	1 \\ 
	2 
	\end{Bmatrix}+l+2\kappa$, $\kappa\in \mathbb{Z}^{+}$. In the zoomed in limit, $\omega_m \sim 2\kappa=L k$ dominates the sum over modes, and therefore 
	\begin{equation}
	\sum_{\kappa} \to \frac{L}{2} \int dk~~~.
	\end{equation}
	We write eq.\eqref{cr2} by substituting $\omega_m=kL$ and replacing sum over modes by integration over $k$ at $\mathcal{O}(q)$
	\begin{equation}
	\begin{split}
	\sqrt{2\omega_{\vec{p}}}~\widetilde{a}^{\dagger}_{\omega_{\vec{p}}}
	&=\widetilde{\mathfrak{C}}~L \int d\Delta_{\vec{p}} ~e^{-i \Delta_{\vec{p}}L\Big[ \frac{\pi}{2}+\frac{i}{2}\log\Big(\frac{\Delta_{\vec{p}}+m}{\Delta_{\vec{p}}-m}\Big)\Big]}~ e^{i\omega_{\vec{p}}L\Big[\frac{\pi}{2}+\frac{i}{2}\log\Big(\frac{\omega_{\vec{p}}+m}{\omega_{\vec{p}}-m}\Big)\Bigg]}\\
	&~~~~~~~~\frac{2\pi}{L} \sum_{m}^{}\frac{~ q~\hat{\chi}_m}{\omega_m}\Bigg(\delta\Big[ (\omega_{\vec{p}}-\Delta_{\vec{p}})+\frac{\omega_m}{L}\Big]-\delta(\omega_{\vec{p}}-\Delta_{\vec{p}})\Bigg)\sqrt{2\Delta_{\vec{p}}} ~\frac{a^{\dagger}_{\Delta_{\vec{p}}}}{\mathcal{C}}\\
	&=\widetilde{\mathfrak{C}}~L \frac{2\pi q}{L} ~\frac{L}{2}\int_{0}^{\infty} dk \frac{1}{L~k}\int_{-\infty}^{\infty} d\Delta_{\vec{p}}~~ e^{-i \Delta_{\vec{p}}L\Big[ \frac{\pi}{2}+\frac{i}{2}\log\Big(\frac{\Delta_{\vec{p}}+m}{\Delta_{\vec{p}}-m}\Big)\Big]}~ e^{i\omega_{\vec{p}}L\Big[\frac{\pi}{2}+\frac{i}{2}\log\Big(\frac{\omega_{\vec{p}}+m}{\omega_{\vec{p}}-m}\Big)\Big]}\\
	&~~~~~~~~~~~~~~~~~~~~~~~~~\Bigg(\delta\Big[ (\omega_{\vec{p}}-\Delta_{\vec{p}})+k\Big]-\delta(\omega_{\vec{p}}-\Delta_{\vec{p}})\Bigg)\hat{\chi}_k\sqrt{2\Delta_{\vec{p}}} ~\frac{a^{\dagger}_{\Delta_{\vec{p}}}}{\mathcal{C}}~~~.
	\end{split}
	\end{equation}
	Now, we can perform the $\Delta_{\vec{p}}$ integral to simplify the expression and finally the dreesed creation mode is given by  
	\begin{equation}
	\label{mixedreptt}
	\begin{split}
	\sqrt{2\omega_{\vec{p}}}~\widetilde{a}^{\dagger}_{\omega_{\vec{p}}}
	&=\widetilde{\mathfrak{C}}~\pi~q~  \int_{0}^{\infty} dk \frac{1}{k} \Bigg(e^{-i (\omega_{\vec{p}}+k)L\Big[ \frac{\pi}{2}+\frac{i}{2}\log\Big(\frac{\omega_{\vec{p}}+k+m}{\omega_{\vec{p}}+k-m}\Big)\Big]}~ e^{i\omega_{\vec{p}}L\Big[\frac{\pi}{2}+\frac{i}{2}\log\Big(\frac{\omega_{\vec{p}}+m}{\omega_{\vec{p}}-m}\Big)\Big]}\\
	&~~~~~~~~\hat{\chi}_k\sqrt{2(\omega_{\vec{p}}+k)}
	\frac{a^{\dagger}_{\omega_{\vec{p}}+k}}{\mathcal{C}}
	-\sqrt{2\omega_{\vec{p}}}~\hat{\chi}_k
	~\frac{a^{\dagger}_{\omega_{\vec{p}}}}{\mathcal{C}}\Bigg)~~~.
	\end{split}
	\end{equation}
	The dressed creation mode acting on the vacuum $\ket{0}$ is the Faddeev-Kulish dressed state. In this way of writing the expression, the frequency of the creation mode of the scalar field will get shifted from $\omega_{\vec{p}}$ to  $\omega_{\vec{p}}+k$. The external leg of the scalar field will get dressed due to the soft Wilson line and this has a nice pictorial representation in terms of the frequency of the scalar field mode, see fig.\ref{fig:mixedrep}. After further reexpressing the CFT operator in terms of the scalar field mode and perform the frequency integral $\Delta_{\vec{p}}$, we get this representation. We refer eq.\eqref{mixedreptt} as ``mixed representation'' (mixed between the ``CFT representation'' and the ``flat space representation'') since the Wilson line operator still contain the CFT current operator.\footnote{For understanding the soft modes of photon in a pictorial way, it's better to express the expression in the mixed representation like in eq.\eqref{mixedreptt}. In terms of CFT operators, the soft modes will correspond to the CFT current operator dual to the gauge field.}    
	
	\section{AdS correction to the Faddeev-Kulish dressed state}
	\label{adscorr}
	In this section, we will explore AdS radius correction to the Faddeev-Kulish dressed state. 
	
	From the previous subsection \ref{mixed}~(ref.eq.\eqref{cr2}), we note the expression for the dressed creation mode written in terms of the undressed mode 
	\begin{equation}
	\begin{split}
	\sqrt{2\omega_{\vec{p}}}~\widetilde{a}^{\dagger}_{\omega_{\vec{p}}}
	&=\widetilde{\mathfrak{C}}~L \int d\Delta_{\vec{p}} ~e^{-i \Delta_{\vec{p}}L\Big[ \frac{\pi}{2}+\frac{i}{2}\log\Big(\frac{\Delta_{\vec{p}}+m}{\Delta_{\vec{p}}-m}\Big)\Big]}~ e^{i\omega_{\vec{p}}L\Big[\frac{\pi}{2}+\frac{i}{2}\log\Big(\frac{\omega_{\vec{p}}+m}{\omega_{\vec{p}}-m}\Big)\Big]}\\
	&~~~~~~~~\int d\tau ~e^{-iL \tau(\omega_{\vec{p}}-\Delta_{\vec{p}})}~e^{iq \int_{\Gamma(\tau,\hat{p})}j_a dx^a}~\sqrt{2\Delta_{\vec{p}}} ~\frac{a^{\dagger}_{\Delta_{\vec{p}}}}{\mathcal{C}}~~~.
	\end{split}
	\end{equation}
	
	Now, in this expression the CFT current operator appears in the Wilson line which makes the formula to be in the mixed representation. We map the CFT current operators to creation/annhilation modes of the photon to express our AdS correction to the Faddeev-Kulish dressed state. Now, we have to write the CFT current operator appeared in the expression for the Wilson line 
	$$e^{iq\int_0^{\tau}j_{\tau^{\prime}}d\tau^{\prime}}=e^{iq\int_0^{\tau}\left(j^+_{\tau^{\prime}}+j^-_{\tau^{\prime}}\right)d\tau^{\prime}}$$
	in terms of the photon creation/annihilation operatos. In the expression of $j_{\tau^{\prime}}$, we use $j^+_{\tau^{\prime}}$ and $j^-_{\tau^{\prime}}$ to denote the positive and negative frequency modes of the photon.
	
	Here, we choose a particular path for the Wilson line. The path is such that the global time coordinate varies from $0$ to $\tau$ and the angular direction remains unchanged. To emphasize, we can do this simplification since we are interested in the soft Wilson line dressing to dress the field and in that soft limit the Wilson line is independent of the path because the field strength term can be ignored. Now, in next section \ref{ADSC}, we explore the AdS corrected modes of the photon in terms of CFT current operators. Then, we invert this mapping in the next to next section \ref{INV}, to express the CFT current operators to AdS corrected modes of the photon. 
	
	\subsection{AdS corrected photon modes in terms of CFT current operators}
	\label{ADSC}
	Free photon fields in flat spacetime can be mode expanded as 
	\begin{equation}
	{ A}_{\mu}(x)=\int {d^3 \vec{q}\over (2\pi)^3} {1\over{\sqrt{2\omega_{\vec{q}}}}} \sum_{\lambda=\pm}\left(
	\varepsilon^{(\lambda)}_{\mu} \hat{a}^{(\lambda)}_{\vec{q}} \, e^{i q\cdot x}
	+
	\varepsilon^{(\lambda)*}_{\mu} \hat{a}^{(\lambda)\dagger}_{\vec{q}} \, e^{-i q\cdot x}
	\right)~~~,
	\end{equation}
	where, $\varepsilon^{(\lambda)}_{\mu}$ are polarization vectors. The creation/annihilation modes of the photon are given by  
	\begin{equation}
	\begin{split}
	\sqrt{2\omega_{\vec{q}}}~\mathbf{a}^{(\lambda)\dagger}_{\vec{q}} =&-i \int d^3 \vec{x} \, \varepsilon^{(\lambda),\mu} e^{i q\cdot x} \overleftrightarrow{\partial_0} {{ A}}_{\mu}(x)~~~ \\
	\sqrt{2\omega_{\vec{q}}}~\mathbf{a}_{\vec{q}}^{(\lambda)} =& ~i \int d^3 \vec{x} \,  (\varepsilon^{(\lambda),\mu})^*  e^{-i q\cdot x} \overleftrightarrow{\partial_0} {{ A}}_{\mu}(x)~~~ . \\
	\end{split}
	\end{equation}
	Now, we express the creation/annihilation modes of the photon in terms of the CFT current operators. We consider the magnetic boundary condition which implies the the gauge field is related to a the bundary CFT current as
	\begin{equation}
	{A}_{\mu}(\rho,x)\xrightarrow[\rho \to \frac \pi 2]{} \cos\rho \, j_{\mu}(x)~~~ .
	\end{equation}
	
	Now, implementing HKLL bulk reconstruction, we have the free photon felds in AdS expressed in terms of the CFT current operator as 
	\begin{equation}
	\label{bulkphoton}
	{A}_{\mu}(\rho,x)=\frac{1}{\pi}\int_{\mathcal{T}}d\tau'~~ \int d^2\Omega' ~	\left[
	K^V_{\mu}(\rho,x;x') \epsilon_{\tau' a b}\partial^{a}j^{b}  (x')
	+
	K^S_{\mu}(\rho,x;x')  \partial_{a}j^{a}(x')
	\right]+\text{h.c.}
	~~~.
	\end{equation}
	Here, the $a, b$ indices denotes the $S^2$ coordinates. The photon field kernels $K^V$ and $K^S$ are obtained by vector harmonic decomposition of the photon field using the treatment of Wald and Ishibashi\cite{Ishibashi:2004wx}. The photon field kernels are worked out in great detail in \cite{Hijano:2020szl, Banerjee:2022oll}.
	
	In \cite{Banerjee:2022oll}, the authors have studied the $1/L^2$ AdS correction to the photon field kernels. Using that, the AdS correction to the photon field modes is derived. The result is given by 
	\begin{equation} 
	\label{exp}
	\small 
	\begin{split}
	\sqrt{2\omega_{\vec{q}}}~\mathbf{a}_{\vec{q}}^{\text{AdS}\, (-)} &=  i \int r^2 dr \, \int d\Omega \, \int\limits_{0}^{\pi} d\tau' \int d\Omega' \int d\omega \frac{1 + z \bar{z}}{\sqrt{2}} \sum_{l,m,l',m'} j_{l'}(r \omega_{\vec{q}}) j_{l}(r \omega) \left(\frac{l(l+1)}{4 \omega^2 L^2}\right)\\
	& ~\times  \Bigg\{ i (\omega + \omega_{\vec{q}})\frac{Y_{lm}\left(\Omega'\right)}{l(l+1)} Y_{l'm'}(\Omega) Y^*_{l'm'}(\Omega_{q}) \partial_{\bar{z}}Y^*_{lm}\left(\Omega \right) (- i)^{- l + l'} e^{-i (\omega - \omega_{\vec{q}}) t} e^{i \omega L \left( \tau' - \frac{\pi}{2}\right)} D^{\bar{z}'} j^-_{\bar{z}'}\\
	&-i (\omega - \omega_{\vec{q}}) \frac{Y^*_{lm}\left(\Omega'\right)}{l(l+1)} Y_{l'm'}(\Omega) Y^*_{l'm'}(\Omega_{q}) \partial_{\bar{z}}Y_{lm}\left(\Omega \right) (-i)^{l'} (i)^{- l} e^{i (\omega + \omega_{\vec{q}}) t} e^{-i \omega L \left( \tau' - \frac{\pi}{2}\right)}\, D^{\bar{z}'} j^+_{\bar{z}'}\Bigg\}~~~.
	\end{split}
	\end{equation}
	
	We can further simplify the expression in eq.\eqref{exp} by performing the derivatives of the spherical harmonics on the $S^2$. We make use of the orthonormality and completeness of the spherical harmonics and orthonormality of the spherical bessel function to do further simplification
	\begin{equation}
	\begin{split}
	\int d\Omega Y_{lm}\left(\Omega\right) Y^*_{l'm'}\left(\Omega\right) &= \delta_{ll'} \delta_{mm'} \label{ylm.int}\\
	\sum_{l,m} Y_{lm}\left(\Omega_{q}\right)Y^*_{lm}\left(\Omega\right) &= \delta^{(2)} \left(\Omega_{q},\Omega\right) \,\\
	\int\limits_{0}^{\infty} r^2 dr j_{l}(r \omega) j_{l}(r \omega_{\vec{q}}) &= \frac{\pi}{2 \omega^2_{\vec{q}}} \delta(\omega - \omega_{\vec{q}}) ~~~.
	\end{split}
	\end{equation}
	
	After simplification, we have the expression of the annihilation operator of an outgoing photon of negative helicity given by  
		\begin{equation}
		\begin{split}
		&\sqrt{2\omega_{\vec{q}}}~\mathbf a^{\text{AdS} (-)}_{\vec{q}}\\
		&=\frac{1}{32\pi\omega_{\vec{q}}^2L^2 }\frac{1+z_q\bar{z}_q}{\sqrt{2}\omega_{\vec{q}}}\int d\tau^{\prime}~e^{-i\omega_{\vec{q}}L\left(\frac{\pi}{2}-\tau^{\prime}\right)}\int d^2z^{\prime} \int d^2z_w
		\Bigg[\frac{(1+z^{\prime}\bar{z}^{\prime})^2(1+z_w\bar{z}_w)^2}{(\bar{z}_w-\bar{z}^{\prime})^2(z_q-z_w)^3}\Bigg]\\
		&~~~~~~~~\partial_{z^{\prime}}j^-_{\bar{z}^{\prime}}(\tau^{\prime},z^{\prime},\bar{z}^{\prime})~.
		\end{split}
		\end{equation}   
	The $1/L^2$ corrected mode is expressed in terms of a CFT current operator smeared over the boundary $S^2$ and we denote by $\mathbf a_{\vec{q}}^{\text{AdS}(-)}$. This mode is physically a bit different from the flat space mode of the photon obtained in the zoomed in limit. The difference is that here an extra $z_w$ integral correponding to integration over intermediate angles $(z_w, \bar{z}_w)$.

	\subsection{Inverse mapping: CFT current operators mapped to AdS corrected photon modes}
	\label{INV}
	In this section, we evaluate the inverse mapping. That means, we express the CFT current operators in terms of the AdS corrected creation/annihilation modes of the photon.
	
	From the previous section, we note that the annihilation operator of an outgoing photon of negative helicity is given by 
		\begin{equation}\label{corrected}
		\begin{split}
		&\sqrt{2\omega_{\vec{q}}}~\mathbf a^{\text{AdS} (-)}_{\vec{q}}\\
		&=\frac{1}{32\pi\omega_{\vec{q}}^2L^2 }\frac{1+z_q\bar{z}_q}{\sqrt{2}\omega_{\vec{q}}}\int d\tau^{\prime}~e^{-i\omega_{\vec{q}}L\left(\frac{\pi}{2}-\tau^{\prime}\right)}\int d^2z^{\prime} \int d^2z_w
		\Bigg[\frac{(1+z^{\prime}\bar{z}^{\prime})^2(1+z_w\bar{z}_w)^2}{(\bar{z}_w-\bar{z}^{\prime})^2(z_q-z_w)^3}\Bigg]\\
		&~~~~~~~~\partial_{z^{\prime}}j^-_{\bar{z}^{\prime}}(\tau^{\prime},z^{\prime},\bar{z}^{\prime})~.
		\end{split}
		\end{equation}

	Acting with a $\partial_{\bar{z}_q}$ on both sides of the eq.\eqref{corrected} we have
	\begin{equation}
	\begin{split}
	\partial_{\bar{z}_q}\Bigg(\frac{64\pi \omega_{\vec{q}}^{7/2}L^2}{1+z_{q}\bar{z}_{q}}
	\mathbf a^{\text{AdS} (-)}_{\vec{q}}\Bigg)&=\int d\tau^{\prime}~e^{-i\omega_{\vec{q}}L\left(\frac{\pi}{2}-\tau^{\prime}\right)}\int d^2z^{\prime}~\int d^2z_w~\partial_{\bar{z}_q}\Bigg[\frac{(1+z^{\prime}\bar{z}^{\prime})^2(1+z_w\bar{z}_w)^2}{(\bar{z}_w-\bar{z}^{\prime})^2(z_q-z_w)^3}\Bigg]\\
	&~~~~\partial_{z^{\prime}}j^{-}_{\bar{z}^{\prime}}\left(\tau^{\prime},z^{\prime},\bar{z}^{\prime}\right)~~~.
	\end{split}
	\end{equation}
	Now, using the identity 
	\begin{equation}
	\begin{split}
	\partial_{\bar{z}_q} \frac{1}{(z_q-z_w)^3}&=(2\pi)\frac{(-1)^2}{2!}\partial_{z_q}^2\delta^{(2)}(z_q,z_w)\\
	&=(2\pi)\frac{(-1)^2}{2!}\partial_{z_w}^2\delta^{(2)}(z_q,z_w)~~~,
	\end{split}
	\end{equation}
	we have 
	\begin{equation} 
	\begin{split}
	\partial_{\bar{z}_q}\Bigg(\frac{64\pi \omega_{\vec{q}}^{7/2}L^2}{1+z_{q}\bar{z}_{q}}
	\mathbf a^{\text{AdS} (-)}_{\vec{q}}\Bigg)&=\int d\tau^{\prime}~e^{-i\omega_{\vec{q}}L\left(\frac{\pi}{2}-\tau^{\prime}\right)}\int d^2z^{\prime}~\int d^2z_w~\pi~\partial_{z_w}^2\delta^{(2)}(z_q,z_w)\\
	&~~~~~\times \Bigg[\frac{(1+z^{\prime}\bar{z}^{\prime})^2(1+z_w\bar{z}_w)^2}{(\bar{z}_w-\bar{z}^{\prime})^2}\Bigg]\partial_{z^{\prime}}j^{-}_{\bar{z}^{\prime}}\left(\tau^{\prime},z^{\prime},\bar{z}^{\prime}\right)~~~.
	\end{split}
	\end{equation}
	
	After performing the $d^2z_w$ integral using the delta function $\delta^{(2)}(z_q,z_w)$ to simplify the expression a bit. The steps of the detailed calculation we save in the appendix \ref{simple}, section \ref{STEPS}. After simplification the expression we get is  
	\begin{equation}
	\label{smp1}
	\begin{split}
	\partial_{\bar{z}_q}\Bigg(\frac{64 \omega_{\vec{q}}^{7/2}L^2}{1+z_{q}\bar{z}_{q}}
	\mathbf a^{\text{AdS} (-)}_{\vec{q}}\Bigg)&=\int d\tau^{\prime}~e^{-i\omega_{\vec{q}}L\left(\frac{\pi}{2}-\tau^{\prime}\right)}\int d^2z^{\prime}~(1+z^{\prime}\bar{z}^{\prime})^2\frac{2\bar{z}_q^2}{(\bar{z}_q-\bar{z}^{\prime})^2}~~\partial_{z^{\prime}}j^{-}_{\bar{z}^{\prime}}\left(\tau^{\prime},z^{\prime},\bar{z}^{\prime}\right)~~~.
	\end{split}
	\end{equation}
	Now, further acting $\partial_{z_q}$ on the simplified expression eq.\eqref{smp1} 
	we finally get 
	\begin{equation}
	\begin{split}
	\partial_{z_q}\Bigg[\partial_{\bar{z}_q}\Bigg(\frac{64 \omega_{\vec{q}}^{7/2}L^2}{1+z_{q}\bar{z}_{q}}
	\mathbf a^{\text{AdS} (-)}_{\vec{q}}\Bigg)\Bigg]&=\int d\tau^{\prime}~e^{-i\omega_{\vec{q}}L\left(\frac{\pi}{2}-\tau^{\prime}\right)}\Big(-8\pi z_q( 1+z_q\bar{z}_q)^2\Big)\partial_{z_q}j^{-}_{\bar{z}_q}\left(\tau^{\prime},z_q,\bar{z}_q\right)~~~.
	\end{split}
	\end{equation} 
	The details is saved in the appendix \ref{simple}, section \ref{STEPS}. 
	
	The inverse mapping of the CFT current operator for the negative frequency in terms of the AdS corrected photon annihilation mode is given by 
	\begin{equation}
	\small 
	\label{zq}
	\partial_{z_q}j^{-}_{\bar{z}_q}\left(\tau^{\prime},z_q,\bar{z}_q\right)=\frac{L}{(2\pi)^2}\int d\omega_{\vec{q}}~e^{-i\omega_{\vec{q}}L\left(\tau^{\prime}-\frac{\pi}{2}\right)}~~\frac{1}{-4z_q(1+z_q\bar{z}_q)^2}\partial_{z_q}\Bigg[\partial_{\bar{z}_q}\Bigg(\frac{64 \omega_{\vec{q}}^{7/2}L^2}{1+z_{q}\bar{z}_{q}}
	\mathbf a^{\text{AdS} (-)}_{\vec{q}}\Bigg)\Bigg]~~~.
	\end{equation} 
	
	Now, we multiply both sides by $\frac{1}{\bar{z}_q-\bar{z}^{\prime}}$ and integrate with respect to $d^2z_q$ and obtain
	\begin{equation}
	\small 
	j^{-}_{\bar{z}^{\prime}}(\tau^{\prime},z^{\prime},\bar{z}^{\prime})=\frac{-L}{(2\pi)^3}\int d\omega_{\vec{q}}~e^{-i\omega_{\vec{q}}L\left(\tau^{\prime}-\frac{\pi}{2}\right)}\int d^2z_q\frac{1}{\bar{z}_q-\bar{z}^{\prime}}\frac{1}{-4z_q(1+z_q\bar{z}_q)^2}\partial_{z_q}\Bigg[\partial_{\bar{z}_q}\Bigg(\frac{64 \omega_{\vec{q}}^{7/2}L^2}{1+z_{q}\bar{z}_{q}}
	\mathbf a^{\text{AdS} (-)}_{\vec{q}}\Bigg)\Bigg]~~~.
	\end{equation}
	
	Now, we can express eq.\eqref{zq} in terms of the coordinates $z^{\prime}$ and $\bar{z}^{\prime}$
	\begin{equation}
	\small 
	\partial_{z^{\prime}}j^{-}_{\bar{z}^{\prime}}\left(\tau^{\prime},z^{\prime},\bar{z}^{\prime}\right)=\frac{L}{(2\pi)^2}\int d\omega_{\vec{q}}~e^{-i\omega_{\vec{q}}L\left(\tau^{\prime}-\frac{\pi}{2}\right)}~~\frac{1}{-4z^{\prime}(1+z^{\prime}\bar{z}^{\prime})^2}\partial_{z_q}\Bigg[\partial_{\bar{z}_q}\Bigg(\frac{64 \omega_{\vec{q}}^{7/2}L^2}{1+z_{q}\bar{z}_{q}}
	\mathbf a^{\text{AdS} (-)}_{\vec{q}}\Bigg)\Bigg]\Bigg|_{(z_q,\bar{z}_q)=(z^{\prime},\bar{z}^{\prime})}~~~.
	\end{equation} 
	
	Similarly, we have 
	\begin{equation}
	\small 
	\begin{split}
	\partial_{\bar{z}^{\prime}}j^{-}_{z^{\prime}}\left(\tau^{\prime},z^{\prime},\bar{z}^{\prime}\right)&=\frac{L}{(2\pi)^2}\int d\omega_{\vec{q}}~e^{-i\omega_{\vec{q}}L\left(\tau^{\prime}-\frac{\pi}{2}\right)}~~\frac{1}{-4\bar{z}^{\prime}(1+z^{\prime}\bar{z}^{\prime})^2}\partial_{\bar{z}_q}\Bigg[\partial_{z_q}\Bigg(\frac{64 \omega_{\vec{q}}^{7/2}L^2}{1+z_{q}\bar{z}_{q}}
	\mathbf a^{\text{AdS} (+)}_{\vec{q}}\Bigg)\Bigg]\Bigg|_{(z_q,\bar{z}_q)=(z^{\prime},\bar{z}^{\prime})}\\
	j^{-}_{z^{\prime}}(\tau^{\prime},z^{\prime},\bar{z}^{\prime}&)=\frac{-L}{(2\pi)^3}\int d\omega_{\vec{q}}~e^{-i\omega_{\vec{q}}L\left(\tau^{\prime}-\frac{\pi}{2}\right)}\int d^2z_q\frac{1}{z_q-z^{\prime}}\frac{1}{-4\bar{z}_q(1+z_q\bar{z}_q)^2}\partial_{\bar{z}_q}\Bigg[\partial_{z_q}\Bigg(\frac{64 \omega_{\vec{q}}^{7/2}L^2}{1+z_{q}\bar{z}_{q}}
	\mathbf a^{\text{AdS} (+)}_{\vec{q}}\Bigg)\Bigg]~~~.
	\end{split}
	\end{equation}
	For the creation mode of the photon we have CFT current operators $j^{+}_{\bar{z}^{\prime}}$ and $j^{+}_{z^{\prime}}$. Here, the $+$ in the superscript of $j$'s denotes the positive frequency mode of the photon. We have the following expressions for the CFT current operators 
	\begin{equation}
	\small 
	\begin{split}
	\partial_{z^{\prime}}j^{+}_{\bar{z}^{\prime}}\left(\tau^{\prime},z^{\prime},\bar{z}^{\prime}\right)&=\frac{L}{(2\pi)^2}\int d\omega_{\vec{q}}~e^{i\omega_{\vec{q}}L\left(\tau^{\prime}-\frac{\pi}{2}\right)}~~\frac{1}{-4z^{\prime}(1+z^{\prime}\bar{z}^{\prime})^2}\partial_{z_q}\Bigg[\partial_{\bar{z}_q}\Bigg(\frac{64 \omega_{\vec{q}}^{7/2}L^2}{1+z_{q}\bar{z}_{q}}
	\mathbf a^{\dagger \text{AdS} (-)}_{\vec{q}}\Bigg)\Bigg]\Bigg|_{(z_q,\bar{z}_q)=(z^{\prime},\bar{z}^{\prime})}\\
	j^{+}_{\bar{z}^{\prime}}(\tau^{\prime},z^{\prime},\bar{z}^{\prime})&=\frac{-L}{(2\pi)^3}\int d\omega_{\vec{q}}~e^{i\omega_{\vec{q}}L\left(\tau^{\prime}-\frac{\pi}{2}\right)}\int d^2z_q\frac{1}{\bar{z}_q-\bar{z}^{\prime}}\frac{1}{-4z_q(1+z_q\bar{z}_q)^2}\partial_{z_q}\Bigg[\partial_{\bar{z}_q}\Bigg(\frac{64 \omega_{\vec{q}}^{7/2}L^2}{1+z_{q}\bar{z}_{q}}
	\mathbf a^{\dagger \text{AdS} (-)}_{\vec{q}}\Bigg)\Bigg]\\
	\partial_{\bar{z}^{\prime}}j^{+}_{z^{\prime}}\left(\tau^{\prime},z^{\prime},\bar{z}^{\prime}\right)&=\frac{L}{(2\pi)^2}\int d\omega_{\vec{q}}~e^{i\omega_{\vec{q}}L\left(\tau^{\prime}-\frac{\pi}{2}\right)}~~\frac{1}{-4\bar{z}^{\prime}(1+z^{\prime}\bar{z}^{\prime})^2}\partial_{\bar{z}_q}\Bigg[\partial_{z_q}\Bigg(\frac{64 \omega_{\vec{q}}^{7/2}L^2}{1+z_{q}\bar{z}_{q}}
	\mathbf a^{\dagger \text{AdS} (+)}_{\vec{q}}\Bigg)\Bigg]\Bigg|_{(z_q,\bar{z}_q)=(z^{\prime},\bar{z}^{\prime})}\\
	j^{+}_{z^{\prime}}(\tau^{\prime},z^{\prime},\bar{z}^{\prime})&=\frac{-L}{(2\pi)^3}\int d\omega_{\vec{q}}~e^{i\omega_{\vec{q}}L\left(\tau^{\prime}-\frac{\pi}{2}\right)}\int d^2z_q\frac{1}{z_q-z^{\prime}}\frac{1}{-4\bar{z}_q(1+z_q\bar{z}_q)^2}\partial_{\bar{z}_q}\Bigg[\partial_{z_q}\Bigg(\frac{64 \omega_{\vec{q}}^{7/2}L^2}{1+z_{q}\bar{z}_{q}}
	\mathbf a^{\dagger \text{AdS} (+)}_{\vec{q}}\Bigg)\Bigg]~~~.
	\end{split}
	\end{equation}
	\subsubsection{Global time component of the CFT current operator mapped to photon modes}
	\label{tau}
	In this section, we evaluate the global time component of the CFT current operator in terms of the photon creation/annihilation modes. Since, in the regime where $F_{MN}=0$, the Wilson line doesn't depend on path. The path for Wilson line we choose in such a way that the global time coordinate varies from $0$ to $\tau$ and the angular direction remains unchanged.
	The CFT current operator in the Wilson line operator  
	\begin{equation}
		e^{iq\int_0^{\tau}j_{\tau^{\prime}}d\tau^{\prime}}=e^{iq\int_0^{\tau}\left(j^+_{\tau^{\prime}}+j^-_{\tau^{\prime}}\right)d\tau^{\prime}}~~~,
	\end{equation}
	can be expressed in terms of the photon creation/annihilation modes.
	
	We use the current conservation equation in the boundary CFT $\partial_a j^a=0$ to relate the global time component of the CFT current operator $j_{\tau}$ in terms of $z$ and $\bar{z}$ components, $j_{z}$ and $j_{\bar{z}}$.
	
	Using the current conservation equation we get,
	\begin{equation}
	\label{consv}
	\begin{split}
	&\partial_{\tau}j^{\tau}+\partial_{z}j^{z}+\partial_{\bar{z}}j^{\bar{z}}=0\\
	\implies &-\partial_{\tau}j_{\tau}+\partial_{z}(g^{z\bar{z}}j_{\bar{z}})+\partial_{\bar{z}}(g^{\bar{z}z}j_{z})=0\\
	\implies &\partial_{\tau}j_{\tau}=\frac{1}{2}(1+z\bar{z})^2(\partial_{z}j_{\bar{z}}+\partial_{\bar{z}}j_{z})+\bar{z}(1+z\bar{z})j_{\bar{z}}+z(1+z\bar{z})j_{z}~~~.
	\end{split}
	\end{equation}
	
	Here, the $S^2$ part of the boundary metric is $ds^2=\frac{4}{(1+z\bar{z})^2} dz d\bar{z}$ which gives $g_{z\bar{z}}=g_{\bar{z}z}=\frac{2}{(1+z\bar{z})^2}$ $\&$ $g^{z\bar{z}}=g^{\bar{z}z}=\frac{1}{2}(1+z\bar{z})^2$. 
Here, in eq.\eqref{consv} we drop $+$ and $-$ superscript in the $z$ component, $\bar{z}$ component and the global time component of the the CFT current operators.	
Using the current conservation equation, global time component of the CFT current operators corresponding to positive and negative frequency modes $j_{\tau}^{+} ~\text{and}~j_{\tau}^{-}$ are given by 
	\begin{equation}
	\small 
	\begin{split} 
	j_{\tau}^{+}
	&=\int_{0}^{\tau} d\tau^{\prime} \Bigg[\frac{1}{2}(1+z^{\prime}\bar{z}^{\prime})^2(\partial_{z^{\prime}}j^{+}_{\bar{z}^{\prime}}+\partial_{\bar{z}^{\prime}}j^{+}_{z^{\prime}})+\bar{z}^{\prime}(1+z^{\prime}\bar{z}^{\prime})j^{+}_{\bar{z}^{\prime}}+z^{\prime}(1+z^{\prime}\bar{z}^{\prime})j^{+}_{z^{\prime}} \Bigg]\\
	&=\frac{1}{2}(1+z^{\prime}\bar{z}^{\prime})^2\\
	&~~~~~\times \Bigg[\frac{L}{(2\pi)^2}\int d\omega_{\vec{q}}~ \frac{1}{iL\omega_{\vec{q}}}e^{-\frac{1}{2} i\pi L \omega_{\vec{q}}}(e^{iL\tau \omega_{\vec{q}}}-1) \frac{1}{-4z^{\prime}(1+z^{\prime}\bar{z}^{\prime})^2}\partial_{z_q}\Bigg[\partial_{\bar{z}_q}\Bigg(\frac{64 \omega_{\vec{q}}^{7/2}L^2}{1+z_{q}\bar{z}_{q}}
	\mathbf a^{\dagger \text{AdS} (-)}_{\vec{q}}\Bigg)\Bigg]\Bigg|_{(z_q,\bar{z}_q)=(z^{\prime},\bar{z}^{\prime})}\\
	&~~~~~+\frac{L}{(2\pi)^2}\int d\omega_{\vec{q}}~ \frac{1}{iL\omega_{\vec{q}}}e^{-\frac{1}{2} i\pi L \omega_{\vec{q}}}(e^{iL\tau \omega_{\vec{q}}}-1) \frac{1}{-4\bar{z}^{\prime}(1+z^{\prime}\bar{z}^{\prime})^2}\partial_{\bar{z}_q}\Bigg[\partial_{z_q}\Bigg(\frac{64 \omega_{\vec{q}}^{7/2}L^2}{1+z_{q}\bar{z}_{q}}
	\mathbf a^{\dagger \text{AdS} (+)}_{\vec{q}}\Bigg)\Bigg]\Bigg|_{(z_q,\bar{z}_q)=(z^{\prime},\bar{z}^{\prime})}\Bigg]\\
	&+\bar{z}^{\prime}(1+z^{\prime}\bar{z}^{\prime})\\
	&\Bigg[
	-\frac{L}{(2\pi)^3}\int d\omega_{\vec{q}}~\frac{1}{iL\omega_{\vec{q}}}e^{-\frac{1}{2} i\pi L \omega_{\vec{q}}}(e^{iL\tau \omega_{\vec{q}}}-1)\int d^2z_q\frac{1}{\bar{z}_q-\bar{z}^{\prime}}\frac{1}{-4z_q(1+z_q\bar{z}_q)^2}\partial_{z_q}\Bigg[\partial_{\bar{z}_q}\Bigg(\frac{64 \omega_{\vec{q}}^{7/2}L^2}{1+z_{q}\bar{z}_{q}}
	\mathbf a^{\dagger \text{AdS} (-)}_{\vec{q}}\Bigg)\Bigg]\\
	&+z^{\prime}(1+z^{\prime}\bar{z}^{\prime})\\
	&\Bigg[-\frac{L}{(2\pi)^3}\int d\omega_{\vec{q}}~ \frac{1}{iL\omega_{\vec{q}}}e^{-\frac{1}{2} i\pi L \omega_{\vec{q}}}(e^{iL\tau \omega_{\vec{q}}}-1)\int d^2 z_q \frac{1}{z_q-z^{\prime}}\frac{1}{-4z_q(1+z_q\bar{z}_q)^2}\partial_{z_q}\Bigg[\partial_{\bar{z}_q}\Bigg(\frac{64 \omega_{\vec{q}}^{7/2}L^2}{1+z_{q}\bar{z}_{q}}
	\mathbf a^{\dagger \text{AdS} (+)}_{\vec{q}}\Bigg)~\Bigg]~,
	\end{split}
	\end{equation}
	
	\begin{equation}
	\small 
	\begin{split}
	j_{\tau}^{-}
	&=\int_{0}^{\tau} d\tau^{\prime} \Bigg[\frac{1}{2}(1+z^{\prime}\bar{z}^{\prime})^2(\partial_{z^{\prime}}j^{-}_{\bar{z}^{\prime}}+\partial_{\bar{z}^{\prime}}j^{-}_{z^{\prime}})+\bar{z}^{\prime}(1+z^{\prime}\bar{z}^{\prime})j^{-}_{\bar{z}^{\prime}}+z^{\prime}(1+z^{\prime}\bar{z}^{\prime})j^{-}_{z^{\prime}} \Bigg]\\
	&=\frac{1}{2}(1+z^{\prime}\bar{z}^{\prime})^2\\
	&\times \Bigg[\frac{L}{(2\pi)^2}\int d\omega_{\vec{q}}~ \frac{1}{-iL\omega_{\vec{q}}}e^{-\frac{1}{2} i\pi L \omega_{\vec{q}}}(e^{iL\tau \omega_{\vec{q}}}-1) \frac{1}{-4z^{\prime}(1+z^{\prime}\bar{z}^{\prime})^2}\partial_{z_q}\Bigg[\partial_{\bar{z}_q}\Bigg(\frac{64 \omega_{\vec{q}}^{7/2}L^2}{1+z_{q}\bar{z}_{q}}
	\mathbf a^{ \text{AdS} (-)}_{\vec{q}}\Bigg)\Bigg]\Bigg|_{(z_q,\bar{z}_q)=(z^{\prime},\bar{z}^{\prime})}\\
	&+\frac{L}{(2\pi)^2}\int d\omega_{\vec{q}}~ \frac{1}{-iL\omega_{\vec{q}}}e^{-\frac{1}{2} i\pi L \omega_{\vec{q}}}(e^{iL\tau \omega_{\vec{q}}}-1) \frac{1}{-4\bar{z}^{\prime}(1+z^{\prime}\bar{z}^{\prime})^2}\partial_{\bar{z}_q}\Bigg[\partial_{z_q}\Bigg(\frac{64 \omega_{\vec{q}}^{7/2}L^2}{1+z_{q}\bar{z}_{q}}
	\mathbf a^{ \text{AdS} (+)}_{\vec{q}}\Bigg)\Bigg]\Bigg|_{(z_q,\bar{z}_q)=(z^{\prime},\bar{z}^{\prime})}\Bigg]\\
	&+\bar{z}^{\prime}(1+z^{\prime}\bar{z}^{\prime})\\
	&\Bigg[
	-\frac{L}{(2\pi)^3}\int d\omega_{\vec{q}}~\frac{1}{-iL\omega_{\vec{q}}}e^{-\frac{1}{2} i\pi L \omega_{\vec{q}}}(e^{iL\tau \omega_{\vec{q}}}-1)\int d^2z_q\frac{1}{\bar{z}_q-\bar{z}^{\prime}}\frac{1}{-4z_q(1+z_q\bar{z}_q)^2}\partial_{z_q}\Bigg[\partial_{\bar{z}_q}\Bigg(\frac{64 \omega_{\vec{q}}^{7/2}L^2}{1+z_{q}\bar{z}_{q}}
	\mathbf a^{ \text{AdS} (-)}_{\vec{q}}\Bigg)\Bigg]\Bigg]\\
	&+z^{\prime}(1+z^{\prime}\bar{z}^{\prime})\\
	&\Bigg[-\frac{L}{(2\pi)^3}\int d\omega_{\vec{q}}~ \frac{1}{-iL\omega_{\vec{q}}}e^{-\frac{1}{2} i\pi L \omega_{\vec{q}}}(e^{iL\tau \omega_{\vec{q}}}-1)\int d^2 z_q \frac{1}{z_q-z^{\prime}}\frac{1}{-4z_q(1+z_q\bar{z}_q)^2}\partial_{z_q}\Bigg[\partial_{\bar{z}_q}\Bigg(\frac{64 \omega_{\vec{q}}^{7/2}L^2}{1+z_{q}\bar{z}_{q}}
	\mathbf a^{\text{AdS} (+)}_{\vec{q}}\Bigg)\Bigg]\Bigg]~.
	\end{split}
	\end{equation}
	
	In deriving the expression for global time component of the CFT current operators, $j_{\tau}^{+}$ and $j_{\tau}^{-}$, we perform the integration over global time coordinates.
	
	The results are given by
	\begin{equation}
	\begin{split}
	\int_{0}^{\tau} e^{i\omega_{\vec{q}}L(\tau^{\prime}-\frac{\pi}{2})} d\tau^{\prime}&=\frac{1}{iL\omega_{\vec{q}}}e^{-\frac{1}{2} i\pi L \omega_{\vec{q}}}(e^{iL\tau \omega_{\vec{q}}}-1)\\
	\int_{0}^{\tau} e^{-i\omega_{\vec{q}}L(\tau^{\prime}-\frac{\pi}{2})} d\tau^{\prime}&=\frac{1}{-iL\omega_{\vec{q}}}e^{\frac{1}{2} i\pi L \omega_{\vec{q}}}(e^{-iL\tau \omega_{\vec{q}}}-1)~~~.
	\end{split}
	\end{equation}
	
	\subsection{Faddeev-Kulish dressed state}
	Now, we express the dressed creation mode for the massive scalar field in terms of the photon creation/annihilation modes and creation mode of the undressed scalar operator. We substitute the expression for CFT current operators, $j^{\pm}_{\tau}$ as 
	$$e^{iq \int_{\Gamma(\tau,\hat{p})}j_a dx^a}=e^{iq\int_0^{\tau}\left(j^+_{\tau^{\prime}}+j^-_{\tau^{\prime}}\right)d\tau^{\prime}}$$
	in the expression of the dressed creation mode of the massive scalar field 
	\begin{equation}
	\begin{split}
	\sqrt{2\omega_{\vec{p}}}~\widetilde{a}^{\dagger}_{\omega_{\vec{p}}}
	&=\widetilde{\mathfrak{C}}~L \int d\Delta_{\vec{p}} ~e^{-i \Delta_{\vec{p}}L\Big[ \frac{\pi}{2}+\frac{i}{2}\log\Big(\frac{\Delta_{\vec{p}}+m}{\Delta_{\vec{p}}-m}\Big)\Big]}~ e^{i\omega_{\vec{p}}L\Big[\frac{\pi}{2}+\frac{i}{2}\log\Big(\frac{\omega_{\vec{p}}+m}{\omega_{\vec{p}}-m}\Big)\Big]}\\
	&~~~~~~~~\int d\tau ~e^{-iL \tau(\omega_{\vec{p}}-\Delta_{\vec{p}})}~e^{iq \int_{\Gamma(\tau,\hat{p})}j_a dx^a}~\sqrt{2\Delta_{\vec{p}}} ~\frac{a^{\dagger}_{\Delta_{\vec{p}}}}{\mathcal{C}}~~~,
	\end{split}
	\end{equation}	
	and express the AdS radius-corrected dressed operator at $\mathcal{O}(q)$.
	The AdS corrected dressed creation operator at $\mathcal{O}(q)$ is given by
	\begin{equation}
	\small 
	\label{tauintegral}
	\begin{split}
	\sqrt{2\omega_{\vec{p}}}~\widetilde{a}^{\dagger}_{\omega_{\vec{p}}}
	&=\widetilde{\mathfrak{C}}~L \int d\Delta_{\vec{p}} ~e^{-i \Delta_{\vec{p}}L\Big[ \frac{\pi}{2}+\frac{i}{2}\log\Big(\frac{\Delta_{\vec{p}}+m}{\Delta_{\vec{p}}-m}\Big)\Big]}~ e^{i\omega_{\vec{p}}L\Big[\frac{\pi}{2}+\frac{i}{2}\log\Big(\frac{\omega_{\vec{p}}+m}{\omega_{\vec{p}}-m}\Big)\Big]}\\
	&~~~~~~~~~\int d\tau ~e^{-iL \tau(\omega_{\vec{p}}-\Delta_{\vec{p}})}~iq~\Bigg(\int_{0}^{\tau}j^{+}_{\tau}(\tau^{\prime},\hat{p}) d\tau^{\prime}+\int_{0}^{\tau}j^{-}_{\tau}(\tau^{\prime},\hat{p}) d\tau^{\prime}\Bigg)\sqrt{2\Delta_{\vec{p}}} ~\frac{a^{\dagger}_{\Delta_{\vec{p}}}}{\mathcal{C}}\\
	&=\widetilde{\mathfrak{C}}~L \int d\Delta_{\vec{p}} ~e^{-i \Delta_{\vec{p}}L\Big[ \frac{\pi}{2}+\frac{i}{2}\log\Big(\frac{\Delta_{\vec{p}}+m}{\Delta_{\vec{p}}-m}\Big)\Big]}~ e^{i\omega_{\vec{p}}L\Big[\frac{\pi}{2}+\frac{i}{2}\log\Big(\frac{\omega_{\vec{p}}+m}{\omega_{\vec{p}}-m}\Big)\Big]}\int d\tau ~e^{-iL \tau(\omega_{\vec{p}}-\Delta_{\vec{p}})}\\
	&\Bigg[iq \times \Bigg\{\frac{1}{2}(1+z^{\prime}\bar{z}^{\prime})^2 \\ 
	&\times \Bigg[\frac{L}{(2\pi)^2}\int d\omega_{\vec{q}}~ \mathcal{I}(\tau,\omega_{q}) \frac{1}{-4z^{\prime}(1+z^{\prime}\bar{z}^{\prime})^2}\partial_{z_q}\Bigg[\partial_{\bar{z}_q}\Bigg(\frac{64 \omega_{\vec{q}}^{7/2}L^2}{1+z_{q}\bar{z}_{q}}
	\mathbf a^{\dagger \text{AdS} (-)}_{\vec{q}}\Bigg)\Bigg]\Bigg|_{(z_q,\bar{z}_q)=(z^{\prime},\bar{z}^{\prime})}\\
	&~~+\frac{L}{(2\pi)^2}\int d\omega_{\vec{q}}~ \mathcal{I}(\tau,\omega_{q}) \frac{1}{-4\bar{z}^{\prime}(1+z^{\prime}\bar{z}^{\prime})^2}\partial_{\bar{z}_q}\Bigg[\partial_{z_q}\Bigg(\frac{64 \omega_{\vec{q}}^{7/2}L^2}{1+z_{q}\bar{z}_{q}}
	\mathbf a^{\dagger \text{AdS} (+)}_{\vec{q}}\Bigg)\Bigg]\Bigg|_{(z_q,\bar{z}_q)=(z^{\prime},\bar{z}^{\prime})}\Bigg]\Bigg\}\\
	&+\bar{z}^{\prime}(1+z^{\prime}\bar{z}^{\prime})\\
	&\Bigg[
	-\frac{L}{(2\pi)^3}\int d\omega_{\vec{q}}~\mathcal{I}(\tau,\omega_{q})\int d^2z_q\frac{1}{\bar{z}_q-\bar{z}^{\prime}}\frac{1}{-4z_q(1+z_q\bar{z}_q)^2}\partial_{z_q}\Bigg[\partial_{\bar{z}_q}\Bigg(\frac{64 \omega_{\vec{q}}^{7/2}L^2}{1+z_{q}\bar{z}_{q}}
	\mathbf a^{\dagger \text{AdS} (-)}_{\vec{q}}\Bigg)\Bigg]\Bigg]\\
	&+z^{\prime}(1+z^{\prime}\bar{z}^{\prime})\\
	&\Bigg[-\frac{L}{(2\pi)^3}\int d\omega_{\vec{q}}~ \mathcal{I}(\tau,\omega_{q})\int d^2 z_q \frac{1}{z_q-z^{\prime}}\frac{1}{-4\bar{z}_q(1+z_q\bar{z}_q)^2}\partial_{\bar{z}_q}\Bigg[\partial_{z_q}\Bigg(\frac{64 \omega_{\vec{q}}^{7/2}L^2}{1+z_{q}\bar{z}_{q}}
	\mathbf a^{\dagger \text{AdS} (+)}_{\vec{q}}\Bigg)\Bigg]\Bigg]\\
	+&\Bigg(\text{terms for the negative frequency mode involving annihilation modes of photon}~\mathbf a_{\vec{q}}^{\text{AdS}(\pm)}\Bigg)\Bigg]\\
	&~~~~\sqrt{2\Delta_{\vec{p}}} ~ ~\frac{a^{\dagger}_{\Delta_{\vec{p}}}}{\mathcal{C}}~~~,\\
	\end{split}
	\end{equation}
	where, $\mathcal{I}(\tau,\omega_{q})$ is given by
	\begin{equation}
	\mathcal{I}(\tau,\omega_{q})=\frac{1}{iL\omega_{\vec{q}}}e^{-\frac{1}{2} i\pi L \omega_{\vec{q}}}\Big(\frac{1}{iL \omega_{\vec{q}}} (e^{iL\tau \omega_{\vec{q}}}-1)-\tau\Big)~~~.
	\end{equation}
	Now, we note that, when acting on the vacuum state, the set of modes with the negative frequency involves annihilation modes and thus kills the state. In the expression for dressed state, these terms will not contribute. The dressed creation operator $\widetilde{a}^{\dagger}_{\omega_{\vec{p}}}$ acting on the vacuum $\ket{0}$ gives the AdS corrected Faddeev-Kulish dressed state. 
	
	In appendix \ref{simple}, in eq.\eqref{aar} we have written the expression by dropping the terms for the negative frequency mode involving annihilation modes of photon~$\mathbf a_{\vec{q}}^{\text{AdS}(\pm)}$. We can further perform the global time, $\tau$ integral to simplify the expression. After performing the global time, $\tau$ integral the integral evaluates to delta function, using which we can further perform frequency, $\Delta_{\vec{p}}$ integral. We have shifted the final expression to appendix \ref{simple}. After performing the frequency, $\Delta_{\vec{p}}$ integral we have the AdS corrected dressed creation operator in eq.\eqref{final} of appendix \ref{simple}.   
	
	\section{Conclusions and future directions}
	\label{Conclusions}
	In this paper, we have constructed AdS radius correction to the Faddeev-Kulish dressed state. We follow the philosophy of Wilson line dressing in the context of AdS/CFT to arrive at the Faddeev-Kulish dressed state. We study the CFT representation and the mixed representation of the Faddeev-Kulish dressed state. We comment on some speculations to get an IR finite $\mathcal{S}$-matrix from AdS/CFT. We use HKLL bulk reconstruction prescription to construct the soft photon modes in terms of the CFT current operators. Then, after expressing the $1/L^2$ correction to the soft photon modes, we implement AdS radius correction to the Wilson line dressing. We invert the mapping of the AdS radius-corrected soft photon modes in terms of CFT current operators, that means we evaluate the CFT current operators in terms of the AdS radius-corrected modes of the photon. In the Wilson line dressing, we use this inverse mapping between the CFT current operators and soft photon modes to construct the AdS radius-corrected creation mode of the Wilson line dressed scalar field. The dressed mode acting on the vaccuum is the desired Faddeev-Kulish dressed state, which takes into account the AdS radius correction.  
	\subsection*{Future directions.}
	In this section, we list some excellent open questions that would be worth exploring in the future. 
	
	\subsection*{CFT representation of the Faddeev-Kulish dressed state outside the zoomed in limit.}
	It would be interesting to obtain a formula for the Faddeev-Kulish dressed states in the CFT outside the zoomed in limit. For that, we first have to construct the AdS radius-corrected modes of the Wilson line dressed massive scalar field in terms of the dual CFT operator. We will report results of this analysis in an upcoming work.
	\subsection*{Asymptotic charges and states from AdS/CFT.}
	The asymptotic symmetries inevitably lead to the Faddeev-Kulish dressed state since the amplitudes that preserve the asymtotic charge are IR finite, the explicit construction was done in \cite{Choi:2017ylo}. The study of generalizations of asymptotic symmetries in AdS was done in \cite{Compere:2019bua, Compere:2020lrt, Fiorucci:2020xto} using different boundary conditions. An excellent open problem is to apply them to create the physical asymptotic states. It would be fascinating to have the states from AdS/CFT and AdS corrections to it.     
	\subsection*{Faddeev-Kulish state involving soft gravitons from AdS/CFT.} 
	It would be interesting to construct the Faddeev-Kulish asymptotic state involving soft gravitons in the flat limit of AdS/CFT and AdS corrections to it. To do this, first we have to construct the graviton scattering states in terms of the CFT stress tensor operators, then following the Wilson line dressing as studied in \cite{Choi:2019fuq} we can construct the Faddeev-Kulish state. Also, using the mapping between flat space scattering states and CFT Stress tensor operators, we can derive the soft graviton theorem and AdS correction to soft graviton theorem from the $CFT_3$ Stress tensor Ward identities.  
	\subsection*{Principal series decomposition in AdS.}
	In celestial amplitude, generally a principal series decomposition of scattering states are used to translate the IR effects in flat spacetime into celestial amplitudes \cite{Pasterski:2017kqt, Donnay:2018neh, Donnay:2020guq, Duary:2022onm, Kapec:2022xjw}. It would be interesting to study the principal series decomposition in AdS. For that, we have to perform principal series decomposition of the CFT correlation function in terms of the bulk Witten diagram in AdS before taking the flat limit. This would open up a new avenue for investigation into the IR effects in AdS. Moving away from the flat limit, this full fledged AdS principal series decomposition will pave the way for the AdS correction to the Faddeev-Kulish dressing.
	\paragraph{Acknowledgement} 
	I would especially like to thank Eliot Hijano for useful discussions and collaboration in the earlier work. I also thank Joydeep Chakravarty, Sangmin Choi, Nava Gaddam, Hofie Sigridar Hannesdottir, Chethan Krishnan, Priyadarshi Paul, Suvrat Raju, Iain Stewart, Becher Thomas, and definately Daniel Kabat, Alok Laddha, R. Loganayagam, Pronobesh Maity, Jyotirmoy Mukherjee, and Pabitra Ray for correspondence and useful discussions. I gratefully acknowledge support from the grant of the Department of Atomic Energy, Government of India, under project no. RTI4001.  
	
	\appendix
	\newpage 
	\section{AdS corrected Faddeev-Kulish dressed state: some simplifications}
	\label{simple}
	In this appendix \ref{simple}, we express the AdS corrected Faddeev-Kulish dressed state. 
	
	The AdS corrected dressed creation operator at $\mathcal{O}(q)$ is given by
	\begin{equation}
	\label{aar}
	\begin{split}
	\sqrt{2\omega_{\vec{p}}}~\widetilde{a}^{\dagger}_{\omega_{\vec{p}}}
	&=\widetilde{\mathfrak{C}}~L \int d\Delta_{\vec{p}} ~e^{-i \Delta_{\vec{p}}L\Big[ \frac{\pi}{2}+\frac{i}{2}\log\Big(\frac{\Delta_{\vec{p}}+m}{\Delta_{\vec{p}}-m}\Big)\Big]}~ e^{i\omega_{\vec{p}}L\Big[\frac{\pi}{2}+\frac{i}{2}\log\Big(\frac{\omega_{\vec{p}}+m}{\omega_{\vec{p}}-m}\Big)\Big]}\int d\tau ~e^{-iL \tau(\omega_{\vec{p}}-\Delta_{\vec{p}})}\\
	&\Bigg[iq \times \Bigg\{\frac{1}{2}(1+z^{\prime}\bar{z}^{\prime})^2 \\ 
	&\times \Bigg[\frac{L}{(2\pi)^2}\int d\omega_{\vec{q}}~ \mathcal{I}(\tau,\omega_{q}) \frac{1}{-4z^{\prime}(1+z^{\prime}\bar{z}^{\prime})^2}\partial_{z_q}\Bigg[\partial_{\bar{z}_q}\Bigg(\frac{64 \omega_{\vec{q}}^{7/2}L^2}{1+z_{q}\bar{z}_{q}}
	\mathbf a^{\dagger \text{AdS} (-)}_{\vec{q}}\Bigg)\Bigg]\Bigg|_{(z_q,\bar{z}_q)=(z^{\prime},\bar{z}^{\prime})}\\
	&~~+\frac{L}{(2\pi)^2}\int d\omega_{\vec{q}}~ \mathcal{I}(\tau,\omega_{q}) \frac{1}{-4\bar{z}^{\prime}(1+z^{\prime}\bar{z}^{\prime})^2}\partial_{\bar{z}_q}\Bigg[\partial_{z_q}\Bigg(\frac{64 \omega_{\vec{q}}^{7/2}L^2}{1+z_{q}\bar{z}_{q}}
	\mathbf a^{\dagger \text{AdS} (+)}_{\vec{q}}\Bigg)\Bigg]\Bigg|_{(z_q,\bar{z}_q)=(z^{\prime},\bar{z}^{\prime})}\Bigg]\Bigg\}\\
	&+\bar{z}^{\prime}(1+z^{\prime}\bar{z}^{\prime})\\
	&\Bigg[
	-\frac{L}{(2\pi)^3}\int d\omega_{\vec{q}}~\mathcal{I}(\tau,\omega_{q})\int d^2z_q\frac{1}{\bar{z}_q-\bar{z}^{\prime}}\frac{1}{-4z_q(1+z_q\bar{z}_q)^2}\partial_{z_q}\Bigg[\partial_{\bar{z}_q}\Bigg(\frac{64 \omega_{\vec{q}}^{7/2}L^2}{1+z_{q}\bar{z}_{q}}
	\mathbf a^{\dagger \text{AdS} (-)}_{\vec{q}}\Bigg)\Bigg]\Bigg]\\
	&+z^{\prime}(1+z^{\prime}\bar{z}^{\prime})\\
	&\Bigg[-\frac{L}{(2\pi)^3}\int d\omega_{\vec{q}}~ \mathcal{I}(\tau,\omega_{q})\int d^2 z_q \frac{1}{z_q-z^{\prime}}\frac{1}{-4\bar{z}_q(1+z_q\bar{z}_q)^2}\partial_{\bar{z}_q}\Bigg[\partial_{z_q}\Bigg(\frac{64 \omega_{\vec{q}}^{7/2}L^2}{1+z_{q}\bar{z}_{q}}
	\mathbf a^{\dagger \text{AdS} (+)}_{\vec{q}}\Bigg)\Bigg]\Bigg]\Bigg]\\
	&~~~~~~~~~~~~\sqrt{2\Delta_{\vec{p}}} ~ ~\frac{a^{\dagger}_{\Delta_{\vec{p}}}}{\mathcal{C}}~~~,\\
	\end{split}
	\end{equation}
	where, $\mathcal{I}(\tau,\omega_{q})$ is given by
	\begin{equation}
	\mathcal{I}(\tau,\omega_{q})=\frac{1}{iL\omega_{\vec{q}}}e^{-\frac{1}{2} i\pi L \omega_{\vec{q}}}\Big(\frac{1}{iL \omega_{\vec{q}}} (e^{iL\tau \omega_{\vec{q}}}-1)-\tau\Big)~~~.
	\end{equation}
	Now, after performing the global time, $\tau$ integral we have  
	\begin{equation}
	\begin{split}
	&\int d\tau ~e^{-iL \tau(\omega_{\vec{p}}-\Delta_{\vec{p}})} \Big(\frac{1}{iL \omega_{\vec{q}}} (e^{iL\tau \omega_{\vec{q}}}-1)-\tau\Big)\\
	&=\frac{1}{iL \omega_{\vec{q}}} \frac{2\pi}{L} \Bigg(\delta(\omega_{\vec{p}}+\omega_{\vec{q}}-\Delta_{\vec{p}})-\delta(\omega_{\vec{p}}-\Delta_{\vec{p}})\Bigg)-i \frac{2\pi}{L}\frac{\partial}{\partial \Delta_{\vec{p}}}\delta(\omega_{\vec{p}}-\Delta_{\vec{p}})~~~.
	\end{split}
	\end{equation}
	
	After performing the frequency, $\Delta_{\vec{p}}$ integral we have the AdS corrected dressed creation operator 
	\begin{equation}
	\small 
	\label{final}
	\begin{split}
	\sqrt{2\omega_{\vec{p}}}~\widetilde{a}^{\dagger}_{\omega_{\vec{p}}}
	&=\frac{2\pi \widetilde{\mathfrak{C}}}{\mathcal{C}} ~~
	\Bigg[-iq \Bigg\{\frac{L}{2}(1+z^{\prime}\bar{z}^{\prime})^2 \times \\
	& \Bigg[\int d\omega_{\vec{q}}~ \Bigg\{ \frac{1}{iL \omega_{\vec{q}}}\Big(e^{\frac{L}{2}\Big[-i\pi \omega_{\vec{q}}-\omega_{\vec{p}}\log\Big(\frac{\omega_{\vec{p}}+m}{\omega_{\vec{p}}-m}\Big)+(\omega_{\vec{p}}+\omega_{\vec{q}})\log\Big(\frac{\omega_{\vec{p}}+\omega_{\vec{q}}+m}{\omega_{\vec{p}}+\omega_{\vec{q}}-m}\Big)\Big]}\sqrt{2(\omega_{\vec{p}}+\omega_{\vec{q}})}a^{\dagger}_{\omega_{\vec{p}}+\omega_{\vec{q}}}-\sqrt{2\omega_{\vec{p}}}a^{\dagger}_{\omega_{\vec{p}}}\Big)\\
	&+ie^{i\omega_{\vec{p}}L\Big[\frac{\pi}{2}+\frac{i}{2}\log\Big(\frac{\omega_{\vec{p}}+m}{\omega_{\vec{p}}-m}\Big)\Big]}\frac{\partial}{\partial \Delta_{\vec{p}}}\Big(\sqrt{2\Delta_{\vec{p}}}a^{\dagger}_{\Delta_{\vec{p}}}e^{-i \Delta_{\vec{p}}L\Big[ \frac{\pi}{2}+\frac{i}{2}\log\Big(\frac{\Delta_{\vec{p}}+m}{\Delta_{\vec{p}}-m}\Big)\Big]}\Big)\Bigg|_{\Delta_{\vec{p}}=\omega_{\vec{p}}}\Bigg\}\\ 
	&~~~~~~~~~\times -\frac{1}{(2\pi)^2} \frac{1}{iL\omega_{\vec{q}}}  e^{-\frac{1}{2} i\pi L \omega_{\vec{q}}}~\frac{1}{-4z^{\prime}(1+z^{\prime}\bar{z}^{\prime})^2}\partial_{z_q}\Bigg[\partial_{\bar{z}_q}\Bigg(\frac{64 \omega_{\vec{q}}^{7/2}L^2}{1+z_{q}\bar{z}_{q}}
	\mathbf a^{\dagger \text{AdS} (-)}_{\vec{q}}\Bigg)\Bigg]\Bigg|_{(z_q,\bar{z}_q)=(z^{\prime},\bar{z}^{\prime})}\\
	&-\int d\omega_{\vec{q}}~\Bigg\{ \frac{1}{iL \omega_{\vec{q}}}\Big(e^{\frac{L}{2}\Big[-i\pi \omega_{\vec{q}}-\omega_{\vec{p}}\log\Big(\frac{\omega_{\vec{p}}+m}{\omega_{\vec{p}}-m}\Big)+(\omega_{\vec{p}}+\omega_{\vec{q}})\log\Big(\frac{\omega_{\vec{p}}+\omega_{\vec{q}}+m}{\omega_{\vec{p}}+\omega_{\vec{q}}-m}\Big)\Big]}\sqrt{2(\omega_{\vec{p}}+\omega_{\vec{q}})}a^{\dagger}_{\omega_{\vec{p}}+\omega_{\vec{q}}}-\sqrt{2\omega_{\vec{p}}}a^{\dagger}_{\omega_{\vec{p}}}\Big)\\
	&+ie^{i\omega_{\vec{p}}L\Big[\frac{\pi}{2}+\frac{i}{2}\log\Big(\frac{\omega_{\vec{p}}+m}{\omega_{\vec{p}}-m}\Big)\Big]}\frac{\partial}{\partial \Delta_{\vec{p}}}\Big(\sqrt{2\Delta_{\vec{p}}}a^{\dagger}_{\Delta_{\vec{p}}}e^{-i \Delta_{\vec{p}}L\Big[ \frac{\pi}{2}+\frac{i}{2}\log\Big(\frac{\Delta_{\vec{p}}+m}{\Delta_{\vec{p}}-m}\Big)\Big]}\Big)\Bigg|_{\Delta_{\vec{p}}=\omega_{\vec{p}}}\Bigg\}\\
	&~~~~~~~~~\times -\frac{1}{(2\pi)^2}\frac{1}{iL\omega_{\vec{q}}}e^{-\frac{1}{2} i\pi L \omega_{\vec{q}}}\frac{1}{-4\bar{z}^{\prime}(1+z^{\prime}\bar{z}^{\prime})^2}\partial_{\bar{z}_q}\Bigg[\partial_{z_q}\Bigg(\frac{64 \omega_{\vec{q}}^{7/2}L^2}{1+z_{q}\bar{z}_{q}}
	\mathbf a^{\dagger \text{AdS} (+)}_{\vec{q}}\Bigg)\Bigg]\Bigg|_{(z_q,\bar{z}_q)=(z^{\prime},\bar{z}^{\prime})}
	\Bigg]\\
	&+\bar{z}^{\prime}(1+z^{\prime}\bar{z}^{\prime})L\\
	&\Bigg[
	\int d\omega_{\vec{q}}~\Bigg\{ \frac{1}{iL \omega_{\vec{q}}}\Big(e^{\frac{L}{2}\Big[-i\pi \omega_{\vec{q}}-\omega_{\vec{p}}\log\Big(\frac{\omega_{\vec{p}}+m}{\omega_{\vec{p}}-m}\Big)+(\omega_{\vec{p}}+\omega_{\vec{q}})\log\Big(\frac{\omega_{\vec{p}}+\omega_{\vec{q}}+m}{\omega_{\vec{p}}+\omega_{\vec{q}}-m}\Big)\Big]}\sqrt{2(\omega_{\vec{p}}+\omega_{\vec{q}})}a^{\dagger}_{\omega_{\vec{p}}+\omega_{\vec{q}}}-\sqrt{2\omega_{\vec{p}}}a^{\dagger}_{\omega_{\vec{p}}}\Big)\\
	&+ie^{i\omega_{\vec{p}}L\Big[\frac{\pi}{2}+\frac{i}{2}\log\Big(\frac{\omega_{\vec{p}}+m}{\omega_{\vec{p}}-m}\Big)\Big]}\frac{\partial}{\partial \Delta_{\vec{p}}}\Big(\sqrt{2\Delta_{\vec{p}}}a^{\dagger}_{\Delta_{\vec{p}}}e^{-i \Delta_{\vec{p}}L\Big[ \frac{\pi}{2}+\frac{i}{2}\log\Big(\frac{\Delta_{\vec{p}}+m}{\Delta_{\vec{p}}-m}\Big)\Big]}\Big)\Bigg|_{\Delta_{\vec{p}}=\omega_{\vec{p}}}\Bigg\}\\
	&\times \frac{1}{(2\pi)^3}\frac{1}{iL\omega_{\vec{q}}}e^{-\frac{1}{2} i\pi L \omega_{\vec{q}}}\int d^2 z_q\frac{1}{\bar{z}_q-\bar{z}^{\prime}} \frac{1}{-4z_q(1+z_q\bar{z}_q)^2}\partial_{z_q}\Bigg[\partial_{\bar{z}_q}\Bigg(\frac{64 \omega_{\vec{q}}^{7/2}L^2}{1+z_{q}\bar{z}_{q}}
	\mathbf a^{\dagger \text{AdS} (+)}_{\vec{q}}\Bigg)\Bigg]
	\Bigg]\\
	&+z^{\prime}(1+z^{\prime}\bar{z}^{\prime})L\\
	&\Bigg[\int d\omega_{\vec{q}}~\Bigg\{ \frac{1}{iL \omega_{\vec{q}}}\Big(e^{\frac{L}{2}\Big[-i\pi \omega_{\vec{q}}-\omega_{\vec{p}}\log\Big(\frac{\omega_{\vec{p}}+m}{\omega_{\vec{p}}-m}\Big)+(\omega_{\vec{p}}+\omega_{\vec{q}})\log\Big(\frac{\omega_{\vec{p}}+\omega_{\vec{q}}+m}{\omega_{\vec{p}}+\omega_{\vec{q}}-m}\Big)\Big]}\sqrt{2(\omega_{\vec{p}}+\omega_{\vec{q}})}a^{\dagger}_{\omega_{\vec{p}}+\omega_{\vec{q}}}-\sqrt{2\omega_{\vec{p}}}a^{\dagger}_{\omega_{\vec{p}}}\Big)\\
	&+ie^{i\omega_{\vec{p}}L\Big[\frac{\pi}{2}+\frac{i}{2}\log\Big(\frac{\omega_{\vec{p}}+m}{\omega_{\vec{p}}-m}\Big)\Big]}\frac{\partial}{\partial \Delta_{\vec{p}}}\Big(\sqrt{2\Delta_{\vec{p}}}a^{\dagger}_{\Delta_{\vec{p}}}e^{-i \Delta_{\vec{p}}L\Big[ \frac{\pi}{2}+\frac{i}{2}\log\Big(\frac{\Delta_{\vec{p}}+m}{\Delta_{\vec{p}}-m}\Big)\Big]}\Big)\Bigg|_{\Delta_{\vec{p}}=\omega_{\vec{p}}}\Bigg\}\\
	&\times\frac{1}{(2\pi)^3}\frac{1}{iL\omega_{\vec{q}}}e^{-\frac{1}{2} i\pi L \omega_{\vec{q}}}\int d^2 z_q \frac{1}{z_q-z^{\prime}} \frac{1}{-4\bar{z}_q(1+z_q\bar{z}_q)^2}\partial_{\bar{z}_q}\Bigg[\partial_{z_q}\Bigg(\frac{64 \omega_{\vec{q}}^{7/2}L^2}{1+z_{q}\bar{z}_{q}}
	\mathbf a^{\dagger \text{AdS} (+)}_{\vec{q}}\Bigg)~\Bigg]~\Bigg]\Bigg\}\Bigg]\left |0\right>~.\\
	\end{split}
	\end{equation}
	
	\subsection{More detailed steps of the derivation of the inverse mapping}
	\label{STEPS}
	In this section \ref{STEPS}, we provide the detailed steps to derive the inverse mapping between the CFT current operators and the photon modes.
	
	The annihilation operator of an outgoing photon of negative helicity is expressed in terms of an integrated expression of the boundary CFT current operator
	\begin{equation}\label{corrected1}
	\begin{split}
     &\sqrt{2\omega_{\vec{q}}}~\mathbf a^{\text{AdS} (-)}_{\vec{q}}\\
     &=\frac{1}{32\pi\omega_{\vec{q}}^2L^2 }\frac{1+z_q\bar{z}_q}{\sqrt{2}\omega_{\vec{q}}}\int d\tau^{\prime}~e^{-i\omega_{\vec{q}}L\left(\frac{\pi}{2}-\tau^{\prime}\right)}\int d^2z^{\prime} \int d^2z_w
     \Bigg[\frac{(1+z^{\prime}\bar{z}^{\prime})^2(1+z_w\bar{z}_w)^2}{(\bar{z}_w-\bar{z}^{\prime})^2(z_q-z_w)^3}\Bigg]\\
     &~~~~~~~~\partial_{z^{\prime}}j^-_{\bar{z}^{\prime}}(\tau^{\prime},z^{\prime},\bar{z}^{\prime})~.
     \end{split}
	\end{equation}
	
	Now, we act with a $\partial_{\bar{z}_q}$ on both sides of the eq.\eqref{corrected1} and get 
	\begin{equation}
	\begin{split}
	\partial_{\bar{z}_q}\Bigg(\frac{64\pi \omega_{\vec{q}}^{7/2}L^2}{1+z_{q}\bar{z}_{q}}
	\mathbf a^{\text{AdS} (-)}_{\vec{q}}\Bigg)&=\int d\tau^{\prime}~e^{-i\omega_{\vec{q}}L\left(\frac{\pi}{2}-\tau^{\prime}\right)}\int d^2z^{\prime}~\int d^2z_w~\partial_{\bar{z}_q}\Bigg[\frac{(1+z^{\prime}\bar{z}^{\prime})^2(1+z_w\bar{z}_w)^2}{(\bar{z}_w-\bar{z}^{\prime})^2(z_q-z_w)^3}\Bigg]\\
	&~~~~~~~~~~\partial_{z^{\prime}}j^{-}_{\bar{z}^{\prime}}\left(\tau^{\prime},z^{\prime},\bar{z}^{\prime}\right)~~~.
	\end{split}
	\end{equation}
	Now, using the identity 
	\begin{equation}
	\begin{split}
	\partial_{\bar{z}_q} \frac{1}{(z_q-z_w)^3}&=(2\pi)\frac{(-1)^2}{2!}\partial_{z_q}^2\delta^{(2)}(z_q,z_w)\\
	&=(2\pi)\frac{(-1)^2}{2!}\partial_{z_w}^2\delta^{(2)}(z_q,z_w)~~~,
	\end{split}
	\end{equation}
	we simplify the expression and get  
	\begin{equation} 
	\begin{split}
	\partial_{\bar{z}_q}\Bigg(\frac{64\pi \omega_{\vec{q}}^{7/2}L^2}{1+z_{q}\bar{z}_{q}}
	\mathbf a^{\text{AdS} (-)}_{\vec{q}}\Bigg)&=\int d\tau^{\prime}~e^{-i\omega_{\vec{q}}L\left(\frac{\pi}{2}-\tau^{\prime}\right)}\int d^2z^{\prime}~\int d^2z_w~\pi~\partial_{z_w}^2\delta^{(2)}(z_q,z_w)\\
	&~~~~~~~\times \Bigg[\frac{(1+z^{\prime}\bar{z}^{\prime})^2(1+z_w\bar{z}_w)^2}{(\bar{z}_w-\bar{z}^{\prime})^2}\Bigg]\partial_{z^{\prime}}j^{-}_{\bar{z}^{\prime}}\left(\tau^{\prime},z^{\prime},\bar{z}^{\prime}\right)~~~.
	\end{split}
	\end{equation}
	Now, we perform the $d^2z_w$ integral using the delta function $\delta^{(2)}(z_q,z_w)$ to simplify the expression a bit 
	\begin{equation}
	\label{inter}
	\begin{split}
	&\int d^2z_w~ \partial^2_{z_w}\delta^{(2)}(z_q,z_w)\Bigg[\frac{(1+z_w\bar{z}_w)^2}{(\bar{z}_w-\bar{z}^{\prime})^2}\Bigg]\\
	&=\frac{2\bar{z}_q^2}{(\bar{z}_q-\bar{z}^{\prime})^2}~~~,
	\end{split}
	\end{equation}
	
	where, in the intermediate steps of eq.\eqref{inter} we use the product rule for double differentiation and expand each term by the following way 
	\begin{equation} 
	\begin{split}
	\partial_{z_w}^2\frac{(1+z_w \bar{z}_w)^2}{(\bar{z}_w-\bar{z}^{\prime})^2}&=\frac{1}{(\bar{z}_w-\bar{z}^{\prime})^2} \partial_{z_w}^2\Big[(1+z_w \bar{z}_w)^2\Big]+(1+z_w \bar{z}_w)^2~\partial_{z_w}^2\frac{1}{(\bar{z}_w-\bar{z}^{\prime})^2}\\
	&~~~~~+2\partial_{z_w}[(1+z_w \bar{z}_w)^2]~\partial_{z_w}\frac{1}{(\bar{z}_w-\bar{z}^{\prime})^2}\\
	&=\frac{2\bar{z}_w^2}{(\bar{z}_w-\bar{z}^{\prime})^2} -2\pi (1+z_w \bar{z}_w)^2~\partial_{z_w}\partial_{\bar{z}_w}\delta^{(2)}(z_w,z^{\prime})\\
	&~~~~~~~~-4\pi\bar{z}_w(1+z_w \bar{z}_w)~\partial_{\bar{z}_w}\delta^{(2)}(z_w,z^{\prime})~~~.
	\end{split}
	\end{equation}
	After simplification the expression becomes 
	\begin{equation}
	\label{smpa}
	\begin{split}
	\partial_{\bar{z}_q}\Bigg(\frac{64 \omega_{\vec{q}}^{7/2}L^2}{1+z_{q}\bar{z}_{q}}
	\mathbf a^{\text{AdS} (-)}_{\vec{q}}\Bigg)&=\int d\tau^{\prime}~e^{-i\omega_{\vec{q}}L\left(\frac{\pi}{2}-\tau^{\prime}\right)}\int d^2z^{\prime}~(1+z^{\prime}\bar{z}^{\prime})^2\frac{2\bar{z}_q^2}{(\bar{z}_q-\bar{z}^{\prime})^2}~~\partial_{z^{\prime}}j^{-}_{\bar{z}^{\prime}}\left(\tau^{\prime},z^{\prime},\bar{z}^{\prime}\right)~~~.
	\end{split}
	\end{equation}
	Now, further acting $\partial_{z_q}$ on the simplified expression eq.\eqref{smpa} and using the identity   
	\begin{equation}
	\partial_{z_q}\frac{1}{(\bar{z}_q-\bar{z}^{\prime})^2}=2\pi (-1)\partial_{\bar{z}_q}\delta^{(2)}(z_q,z^{\prime})~~~,
	\end{equation}
	we get 
	\begin{equation}
	\begin{split}
	\partial_{z_q}\Bigg[\partial_{\bar{z}_q}\Bigg(\frac{64 \omega_{\vec{q}}^{7/2}L^2}{1+z_{q}\bar{z}_{q}}
	\mathbf a^{\text{AdS} (-)}_{\vec{q}}\Bigg)\Bigg]&=\int d\tau^{\prime}~e^{-i\omega_{\vec{q}}L\left(\frac{\pi}{2}-\tau^{\prime}\right)}\Big(-8\pi z_q( 1+z_q\bar{z}_q)^2\Big)\partial_{z_q}j^{-}_{\bar{z}_q}\left(\tau^{\prime},z_q,\bar{z}_q\right)~~~.
	\end{split}
	\end{equation} 
	Now, we get the inverse mapping of the CFT current operator for the negative frequency in terms of the AdS corrected photon annihilation mode
	\begin{equation}
	\partial_{z_q}j^{-}_{\bar{z}_q}\left(\tau^{\prime},z_q,\bar{z}_q\right)=\frac{L}{(2\pi)^2}\int d\omega_{\vec{q}}~e^{-i\omega_{\vec{q}}L\left(\tau^{\prime}-\frac{\pi}{2}\right)}~~\frac{1}{-4z_q(1+z_q\bar{z}_q)^2}\partial_{z_q}\Bigg[\partial_{\bar{z}_q}\Bigg(\frac{64 \omega_{\vec{q}}^{7/2}L^2}{1+z_{q}\bar{z}_{q}}
	\mathbf a^{\text{AdS} (-)}_{\vec{q}}\Bigg)\Bigg]~~~.
	\end{equation} 
	The useful identities used to derive the inverse mapping,  we list in appendix, section \ref{IDT}.
	\subsubsection{Useful identities used to derive the inverse mapping}
	\label{IDT}
	Here, we list the useful identities used to derive the inverse mapping  
	\begin{equation}
	\begin{split}
	\partial_z^n \delta(z)&=\frac{(-1)^n n!}{z^n} \delta(z)\\
	\partial_z \delta^{}(z,w)&=-\partial_w \delta^{}(z,w)\\
	\int \partial_z^n \delta(z,w) f(z)dz&=(-1)^n \partial_z^n f(z)\Big|_{z=w}\\
	\partial_z \delta^{}(z,w) f(w,\bar{w})&=\partial_z \Big[\delta^{(2)}(z,w)f(w,\bar{w})\Big]\\
	&=\partial_z \Big[\delta^{}(z,w)f(z,\bar{z})\Big]\\
	&=\partial_z \delta^{}(z,w)f(z,\bar{z})+\delta^{}(z,w)\partial_z f(z,\bar{z})\\
	\partial^2_z \delta^{}(z,w) f(w,\bar{w})&=\partial^2_z \Big[\delta^{}(z,w)f(w,\bar{w})\Big]\\
	&=\partial^2_z \Big[\delta^{}(z,w)f(z,\bar{z})\Big]\\
	&=\partial_z^2 \delta^{}(z,w)f(z,\bar{z})+\delta^{}(z,w)\partial_z^2 f(z,\bar{z})+2\partial_z\delta^{}(z,w)\partial_z f(z,\bar{z})~~~~\\
	\partial_{\bar{z}}\frac{1}{(z-w)^{n+1}}&=(2\pi)\frac{(-1)^{n}}{n!}\partial_z^n\delta^{(2)}(z,w)~~~,
	\end{split}
	\end{equation}
	where, the delta function $\delta^{(2)}(z,w)$ in the last equation is a complex plane delta function.  
	\providecommand{\href}[2]{#2}\begingroup\raggedright

\end{document}